\documentclass{jfm}

\usepackage{graphicx}
\usepackage{natbib}
\usepackage{amsmath}
\usepackage{amssymb}
\usepackage{graphicx}
\usepackage{subfigure}
\usepackage{color}

\ifCUPmtlplainloaded \else
  \checkfont{eurm10}
  \iffontfound
    \IfFileExists{upmath.sty}
      {\typeout{^^JFound AMS Euler Roman fonts on the system,
                   using the 'upmath' package.^^J}%
       \usepackage{upmath}}
      {\typeout{^^JFound AMS Euler Roman fonts on the system, but you
                   dont seem to have the}%
       \typeout{'upmath' package installed. JFM.cls can take advantage
                 of these fonts,^^Jif you use 'upmath' package.^^J}%
      }
  \else
  \fi
\fi

\ifCUPmtlplainloaded \else
  \checkfont{msam10}
  \iffontfound
    \IfFileExists{amssymb.sty}
      {\typeout{^^JFound AMS Symbol fonts on the system, using the
                'amssymb' package.^^J}%
       \usepackage{amssymb}%
       \let\le=\leqslant  
       \let\ge=\geqslant  
      }{}
  \fi
\fi

\ifCUPmtlplainloaded \else
  \IfFileExists{amsbsy.sty}
    {\typeout{^^JFound the 'amsbsy' package on the system, using it.^^J}%
     \usepackage{amsbsy}}
    {\providecommand\boldsymbol[1]{\mbox{\boldmath $##1$}}}
\fi

\providecommand\bnabla{\boldsymbol{\nabla}}
\providecommand\bcdot{\boldsymbol{\cdot}}

\newsavebox{\astrutbox}
\sbox{\astrutbox}{\rule[-5pt]{0pt}{20pt}}

\newcommand{\be}[1]{\begin{equation}\label{#1}}
\newcommand{\ee}{\end{equation}}
\newcommand{\ba}[1]{\begin{eqnarray}\label{#1}}
\newcommand{\ea}{\end{eqnarray}}
\newcommand{\rf}[1]{(\ref{#1})}
\newcommand{\nn}{\nonumber}

\newcommand{\diag}{\mbox{\rm diag}\,}

\title[Local instabilities in magnetized rotational  flows]{Local instabilities in  magnetized rotational  flows:
A short-wavelength approach}

\author[O. N. Kirillov, F. Stefani and Y. Fukumoto]%
{O.\ls N.\ns K\ls I\ls R\ls I\ls L\ls L\ls O\ls V$^1$%
  \thanks{Email address for correspondence: o.kirillov@hzdr.de},\ns
F.\ns S\ls T\ls E\ls F\ls A\ls N\ls I$^1$\break
\and Y.\ns F\ls U\ls K\ls U\ls M\ls O\ls T\ls O$^2$}

\affiliation{$^1$Helmholtz-Zentrum Dresden-Rossendorf, P.O. Box 510119, D-01314 Dresden, Germany\\[\affilskip]
$^2$Institute of Mathematics for Industry 744 Motooka, Nishi-ku, Fukuoka 819-0395, Japan}

\pubyear{2010}
\volume{650}
\pagerange{119--126}
\date{?; revised ?; accepted ?. - To be entered by editorial office}
\begin{document}

\maketitle

\begin{abstract}
We perform a local stability analysis of rotational flows in the presence of a
constant vertical magnetic field and an azimuthal magnetic field
with a general radial dependence.
Employing the short-wavelength approximation we develop a unified framework
for the investigation of the standard, the helical, and the azimuthal version
of the magnetorotational instability, as well as of current-driven
kink-type instabilities.
Considering the viscous and resistive setup, our main focus is on the
case of small magnetic Prandtl numbers which applies, e.g., to
liquid metal experiments but also to the colder parts of accretion disks.
We show that the inductionless versions of MRI that were
previously thought to be restricted to comparably steep rotation profiles
extend well to the Keplerian case if only the azimuthal field
slightly deviates from its current-free (in the fluid) profile.
We find an explicit criterion separating the pure azimuthal inductionless magnetorotational instability from the regime where this
instability is mixed with the Tayler instability. We further demonstrate that for particular parameter configurations the azimuthal MRI originates as a result of a dissipation-induced instability of the Chandrasekhar's equipartition solution of ideal magnetohydrodynamics.

\end{abstract}


\section{Introduction}

The interaction of rotational flows and magnetic fields is of fundamental importance
for many  geo- and astrophysical problems \citep{RHK13}. On one hand, rotating
cosmic bodies, such as planets, stars, and galaxies are known to
generate magnetic fields by means of the hydromagnetic dynamo effect.
Magnetic fields, in turn, can destabilize rotating flows that would be otherwise
hydrodynamically stable. This effect is particularly important for accretion disks
in active galactic nuclei, dwarf novae and protoplanetary systems,
where it allows for the tremendous enhancement of
outward directed angular momentum transport that is necessary to explain the
typical mass flow rates onto the respective central objects. Although this
{\it magnetorotational instability} (MRI), as we call it now, had been
discovered already in 1959 by Velikhov \citep{Velikhov59} and then confirmed in 1960 by
Chandrasekhar \citep{Chandra60}, it was left to \cite{BH1991} to point out its relevance
for astrophysical accretion processes. Their seminal paper has
inspired many investigations related to the action of MRI in active
galactic nuclei \citep{KROLIK}, X-ray binaries \citep{DONE},
protoplanetary disks \citep{ARMITAGE}, stars \citep{KW2014}, and even planetary cores \citep{PETITDEMANGE}.

An interesting question concerns the non-trivial interplay of
the hydromagnetic dynamo effect and magnetically triggered flow
instabilities.
For a long time, dynamo research had been focussed on how a
prescribed flow can produce a magnetic field and, to a lesser extent,
on how the self-excitation process saturates when the magnetic field
becomes strong enough to act against the source of its own generation.
Similarly, most of the early MRI studies have assumed some prescribed  magnetic
field, e.g., a purely axial or a purely azimuthal field, to assess its
capability for triggering  instabilities and turbulent
angular momentum transport in the flow.
Nowadays, however, we witness an increasing interest in treating the
dynamo effect and instabilities in magnetized flows in a more
self-consistent manner.
Combining both processes one can ask for the existence of
``self-creating dynamos'' \citep{Fuchs99}, i.e.
dynamos whose magnetic field  triggers, at least partly,
the flow structures that are responsible for its self-excitation.

A paradigm of such an essentially  non-linear dynamo problem is the case of an
accretion disk without any externally applied axial magnetic field.
In this case the magnetic field can only be produced in the disk itself,
very likely by a periodic MRI dynamo process \citep{HERAULT}
or some sort of an $\alpha-\Omega$ dynamo
\citep{BRANDENBURG}, the $\alpha$ part of which relies on the
turbulent flow structure arising due to the MRI.
Such a closed loop of magnetic field self-excitation and MRI has
attracted much attention in the past, though
with many unsolved questions
concerning numerical convergence \citep{FROMANG}, the influence of
disk stratification
\citep{STRATIFICATION}, and the role of boundary conditions for the
magnetic field \citep{KAPILA}.

In problems of that kind, a key role is played by the
so-called magnetic Prandtl number ${\rm Pm}=\nu/\eta$, i.e. the ratio of the
viscosity $\nu$ of the
fluid to its magnetic diffusivity $\eta=({\mu}_0 \sigma)^{-1}$ (with ${\mu}_0$ denoting the
magnetic permeability of free space and $\sigma$ the conductivity).
For the case of an axially applied field,
a first systematic study of the ${\rm Pm}$-dependence
of the growth rate and the angular momentum transport coefficient
was carried out by \cite{LESUR}. The closed-loop MRI-dynamo
process, which seems to work well for ${\rm Pm}=O(1)$
\citep{HERAULT}, is much more intricate for small values of
$\rm Pm$, as
they are typical for the outer parts of accretion disks around
black holes and for protoplanetary disks.
At least for stratified disks, \cite{OISHI}
have recently claimed evidence for a critical magnetic Reynolds number
${\rm Rm}_{\rm crit}$
in the order of 3000
that is mainly independent of ${\rm Pm}$.
A somewhat higher value of ${\rm Rm}_{\rm crit}$ had been found earlier
by \cite{FLEMING} when considering resistivity while
still neglecting viscous effects.

Another paradigm of the interplay of self-excitation
and magnetically triggered instabilities
is the so-called Tayler-Spruit dynamo as proposed by
\cite{SPRUIT}. In this particular
(and controversially discussed) model of stellar magnetic field generation,
the $\Omega$ part of the dynamo process
(to produce toroidal field from poloidal field)
is played, as usual, by the differential rotation, while the
$\alpha$ part (to produce poloidal from toroidal field)
is taken over by the flow structure arising from the kink-type Tayler
instability \citep{TAYLER} that sets in when the toroidal
field acquires a critical strength to overcome stable stratification.

At small values of $\rm Pm$, both dynamo and MRI related problems are
very hard to treat numerically. This has to do with the fact that
both phenomena rely on induction effects which require some finite
magnetic Reynolds number. This number is the ratio of
magnetic field production by the velocity to magnetic field dissipation due to
Joule heating. For a fluid flow with typical size $L$ and typical
velocity $V$ it can be expressed as ${\rm Rm}=\mu_0 \sigma L V$.
The numerical difficulty for small ${\rm Pm}$ problems arises then from the
relation that the hydrodynamic Reynolds number, i.e. $\rm Re= Pm^{-1} Rm$, becomes
very large, so that extremely fine structures have to be resolved.
Furthermore, for MRI problems it is additionally necessary that the
magnetic Lundquist number, which is simply a magnetic Reynolds number based
on the Alfv\'en velocity $v_A$,
i.e. $S=\mu_0 \sigma L v_{A}$, must also be in the order of 1.

A complementary way to study the interaction of rotating flows
and magnetic field at small $\rm {Pm}$ and comparably large
$\rm {Rm}$ is by means of liquid metal experiments. As for the dynamo
problem, quite  a number of experiments have been carried out \citep{ZAMM}.
Up to present, magnetic field self-excitation has been attained in the
liquid sodium experiments in Riga \citep{RIGA}, Karlsruhe \citep{KARLSRUHE},
and Cadarache \citep{VKS}. Closely related to these dynamo experiments,
some groups have also attempted to explore the standard version of
MRI (SMRI), which corresponds to the case that a purely
vertical magnetic field is being applied to the flow \citep{SISAN,NORNBERG}.
Recently, the current-driven, kink-type Tayler instability was
identified in a liquid metal experiment
\citep{SEILMAYER2012}, the findings of which were numerically
confirmed in the framework of an integro-differential equation
approach by \cite{WEBER}.

With view on the peculiarities to do numerics, and experiments,
on the standard version of MRI at low $\rm Pm$,
it came as a big surprise when \cite{Hollerbach2005}
showed that the simultaneous
application  of an axial and an azimuthal magnetic field can
change completely the parameter scaling  for the onset of MRI.
For $B_{\phi}/B_{z} \sim 1$, the helical MRI (HMRI), as we call it now,
was shown to work even in the inductionless limit \citep{P11,KS11},
${\rm Pm}=0$, and to be governed by the Reynolds number
${\rm Re}={\rm Rm} {\rm Pm}^{-1}$ and the Hartmann number
${\rm Ha}=S {\rm Pm}^{-1/2}$, quite in contrast to standard MRI (SMRI)
that was known to be governed by ${\rm Rm}$ and $S$ \citep{JGK01}.

Very soon, however, the enthusiasm about this new inductionless
version of MRI cooled down when \cite{Liu2006} showed that HMRI would only
work for relatively steep rotation profiles, see also \citep{JB2013}.
Using a short-wavelength
approximation, they were able to identify a minimum steepness of the
rotation profile $\Omega(R)$, expressed by the
Rossby number ${\rm Ro}:= R(2\Omega)^{-1} \partial \Omega/
\partial R<{\rm Ro}_{\rm LLL}=2(1{-}\sqrt 2)\approx -0.828$.
This limit, which we will call the {\it lower Liu limit} (LLL)
in the following, implies that for $B_{\phi}(R)\propto 1/R$ the inductionless
HMRI does not extend to the
Keplerian case, characterized by
${\rm Ro}_{\rm Kep}=-3/4$.
Interestingly,  \cite{Liu2006} found also a second threshold of
the Rossby number, which we call the
{\it upper Liu limit} (ULL), at
${\rm Ro}_{\rm ULL}=2(1{+}\sqrt 2)\approx +4.828$. This
second limit, which predicts a magnetic destabilization of
extremely stable flows with strongly increasing angular
frequency, has attained nearly
no attention up to present, but will play an important
role in the present paper.

As for the general relation between HMRI and SMRI,
two apparently contradicting observations have to be mentioned.
On one hand, the numerical results of
\cite{Hollerbach2005} had clearly demonstrated a continuous and
monotonic transition between HMRI and SMRI. On the other
hand, HMRI was identified by \cite{Liu2006} as a weakly destabilized inertial
oscillation, quite in contrast to the SMRI which represents a destabilized
slow magneto-Coriolis wave. Only recently,
this paradox was resolved
by showing that the transition
involves a spectral exceptional point at which the
inertial wave branch coalesces with the branch of the
slow magneto-Coriolis wave \citep{KS10}.

The significance of the LLL, together with a variety of further
predicted parameter dependencies, was experimentally confirmed in the
PROMISE facility, a Taylor-Couette cell working with a low ${\rm Pm}$
liquid metal \citep{PRL,NJP,PRE}. Present experimental work at the same
device \citep{SEILMAYER2013}
aims at the characterization of the azimuthal MRI (AMRI),
a non-axisymmetric ``relative'' of
the axisymmetric HMRI, which dominates at
large ratios of $B_\phi$ to $B_z$ \citep{HTR10}.
However, AMRI as well as inductionless MRI modes with any other
azimuthal wavenumber
(which may be relevant at small values of $B_{\phi}/B_z$),
seem also to be  constrained by the LLL as recently shown in a
unified treatment of all inductionless versions of MRI by
\cite{KS12b}.

Actually, it is this apparent failure of HMRI, and AMRI,
to apply to Keplerian profiles that has prevented a wider
acceptance of those inductionless forms of MRI in the
astrophysical community. Given the close proximity of the
LLL ($\approx-0.83$) and the Keplerian Rossby number ($-0.75$), it
is certainly worthwhile to ask whether any physically sensible
modification would allow HMRI to extend to Keplerian flows.

Quite early, the validity of the LLL for $B_{\phi}(R)\propto 1/R$
had been questioned by \cite{RH07}. For the convective instability, they
found an  extension of the LLL to the Keplerian value
in global simulations when at least one of the
radial boundary conditions was assumed
electrically conducting.
Later, though,  by extending the study to the
absolute instability for the traveling
HMRI waves, the LLL was vindicated even for such
modified electrical boundary conditions by \cite{P11}.
\cite{KS11} made a second attempt by investigating
HMRI for non-zero, but low $S$.
For $B_{\phi}(R)\propto 1/R$ it was found that
the essential HMRI mode extends
from $S=0$ only to a value $S\approx0.618$, and allows for
a maximum Rossby number of ${\rm Ro}\approx-0.802$ which is indeed
slightly above the LLL, yet below the Keplerian value.

A third possibility may arise when considering that saturation of
MRI could lead to modified flow structures with parts of steeper
shear, sandwiched with parts of shallower shear
\citep{UMURHAN10}.

A recent letter \citep{KS13} has suggested another way
of extending
the range of applicability of the inductionless versions of
MRI to Keplerian profiles, and beyond. Rather than relying on
modified electrical boundary conditions, or on
locally \textit{steepened} $\Omega(R)$ profiles, we have evaluated
$B_\phi(R)$ profiles that are \textit{flatter} than $1/R$.
The main physical idea behind this attempt is the following: assume that
in some low-${\rm Pm}$ regions,
characterized by $S<<1$ so that standard MRI is reliably
suppressed, ${\rm Rm}$ is still sufficiently
large for inducing significant
azimuthal magnetic fields, either from a prevalent axial
field $B_z$ or by means
of a  dynamo process without any prescribed   $B_z$.
Note that $B_\phi\propto 1/R$ would only appear
in the extreme case of an isolated axial current, while
the other extreme case, $B_\phi \propto R$, would correspond to the
case of a homogeneous axial current density
in the fluid which is already prone to
the kink-type Tayler instability \citep{SEILMAYER2012}, even
at  ${\rm Re}=0$.

Imagine now a real accretion disk with its
complicated conductivity
distributions in radial and axial direction.
For such real disks a large
variety of intermediate $B_\phi(R)$ profiles between
the extreme cases $\propto 1/R$  and  $\propto R$ is
well conceivable. Instead of going into those details,  one can
ask  which $B_{\phi}(R)$ profiles
could make HMRI a viable mechanism for destabilizing
Keplerian rotation profiles.
By defining an appropriate {\it magnetic  Rossby number} ${\rm Rb}$
we showed that the instability extends well beyond the LLL, even reaching
${\rm Ro}=0$  when going to ${\rm Rb}=-0.5$. 
It should be noted that in this extreme case
of uniform rotation the only available energy source of the instability
is the magnetic field. Going then over into the
region of positive ${\rm Ro}$ in the ${\rm Ro}-{\rm Rb}$ plane,
we found a natural connection with the
ULL which was a somewhat mysterious conundrum up to
present.

The present paper represents a significant extension of the short letter
\citep{KS13}. In the first instance, we will present a detailed
derivation of the
dispersion relation for arbitrary azimuthal modes in
viscous, resistive rotational flows under the influence of
a constant axial and a superposed azimuthal field of arbitrary radial
dependence.
For this purpose, we employ general short-wavelength asymptotic series following
\cite{E81,E87}, \cite{B88}, \cite{L91}, \cite{LH91}, \cite{FV95}, \cite{VF98}, \cite{Fukumoto2003}, and \cite{FL03} as well as a WKB approach of \cite{KS10}.

Second, we will discuss in much more detail the stability map
in the ${\rm Ro}-{\rm Rb}$-plane in the inductionless case of vanishing magnetic Prandtl number. For various limits
we will discuss a number of
strict results concerning the stability threshold and the
growth rates. Special focus will be laid on the
role that is played by the line ${\rm Ro}={\rm Rb}$,
and by the point ${\rm Ro}={\rm Rb}=-2/3$ in particular.

Third, we will elaborate the dependence of the
instability on the azimuthal wavenumber and on the ratio of
the axial and radial wavenumbers and prove that the pattern of instability domains in the case of very small, but finite $\rm Pm$, is
governed by a periodic band structure found in the inductionless limit.

Next, we will establish a connection between dissipation-induced destabilization of Chandrasekhar's equipartition solution (a special solution for which the fluid velocity is parallel to the direction of the magnetic field  and magnetic and kinetic
energies are finite and equal  \citep{Chandra56,CH61}) and the azimuthal MRI, and we will explore the links between the Tayler instability and AMRI.

Last, but not least, we will delineate some possible
astrophysical and experimental consequences of our findings, although a comprehensive
discussions of the corresponding details  must be left for future work.

\section{Mathematical setting}
\subsection{Non-linear equations}
The standard set of non-linear equations of
dissipative incompressible magnetohydrodynamics
consists of the Navier-Stokes equation for
the fluid velocity $\boldsymbol{u}$ and of the induction equation for the magnetic field $\boldsymbol{B}$
\ba{m1}
&\frac{\partial \boldsymbol{u}}{\partial t}+\boldsymbol{u} \bcdot \bnabla \boldsymbol{u}-\frac{1}{\mu_0 \rho}\boldsymbol{B}\bcdot \bnabla\boldsymbol{B} +\frac{1}{\rho} \bnabla P-\nu \bnabla^2 \boldsymbol{u}=0,&\nn \\
&\frac{\partial \boldsymbol{B}}{\partial t}+\boldsymbol{u} \bcdot \bnabla \boldsymbol{B} - \boldsymbol{B} \bcdot \bnabla \boldsymbol{u}- \eta \bnabla^2 \boldsymbol{B}=0,&
\ea
where $P=p+\frac{\boldsymbol{B}^2}{2\mu_0}$ is the total pressure, $p$ is the hydrodynamic pressure, $\rho=const$
the density, $\nu=const$ the kinematic
viscosity,
$\eta=(\mu_0 \sigma)^{-1}$ the magnetic diffusivity, $\sigma$ the conductivity of the fluid,
and $\mu_0$
the magnetic permeability of free space.
Additionally, the mass continuity equation for incompressible flows
and the solenoidal condition for the magnetic induction yield
\be{m3}
\bnabla \bcdot \boldsymbol{u} = 0,\quad  \bnabla \bcdot \boldsymbol{B}=0.
\ee

\subsection{Steady state}
We consider the rotational fluid flow in the gap between the radii
$R_1$ and $R_2>R_1$, with an imposed magnetic field sustained by electrical
currents.
Introducing the cylindrical coordinates $(R, \phi, z)$ we consider
the stability of a steady-state background liquid flow with           
the angular
velocity profile $\Omega(R)$ in helical background magnetic field (a
magnetized Taylor-Couette (TC) flow)
\be{m4} \boldsymbol{u}_0(R)=R\,\Omega(R)\,\boldsymbol{
e}_{\phi},\quad p=p_0(R), \quad \boldsymbol{B}_0(R)=B_{\phi}^0(R)\boldsymbol{
e}_{\phi}+B_z^0 \boldsymbol{ e}_z.
\ee
Note that if the azimuthal component is produced by an axial current $I$ confined to $R<R_1$, then
\be{m4a}
B_{\phi}^0(R)=\frac{\mu_0 I}{2 \pi R}.                                   
\ee

Introducing the hydrodynamic Rossby number $({\rm Ro})$ by means of the relation
\be{hro}
{\rm Ro}:=\frac{R}{2\Omega}\partial_R\Omega,
\ee
we find that the solid body rotation corresponds to ${\rm Ro}=0$,
the Keplerian rotation to ${\rm Ro}=-3/4$, whereas
the velocity profile $\Omega(R)\sim R^{-2}$ corresponds to ${\rm Ro}=-1$. Note that although
Keplerian rotation is not possible globally  in TC-flow, an approximate Keplerian profile can locally be achieved.

Similarly, following \cite{KS13} we define the magnetic Rossby number
\be{mro}
{\rm Rb}:=\frac{R}{2B_{\phi}^0 R^{-1}}\partial_R (B_{\phi}^0 R^{-1}).
\ee
${\rm Rb}=0$ results from a linear dependence of
the magnetic field on the radius, $B_{\phi}^0(R)\propto R$, as
it would be produced by a homogeneous axial current in the fluid.
${\rm Rb}=-1$ corresponds to the radial dependence given by Eq.~\rf{m4a}. Note that the logarithmic derivatives $\alpha$ and $\beta$ introduced in the work by \cite{OP1996} are nothing else but the doubled hydrodynamic and magnetic Rossby numbers: $\alpha=2{\rm Ro}$, $\beta=2{\rm Rb}$.

\subsection{Linearization with respect to non-axisymmetric perturbations}

To describe natural oscillations in the neighborhood of the magnetized Taylor-Couette flow we linearize equations \rf{m1}-\rf{m3} in the vicinity of the stationary solution \rf{m4} assuming general perturbations $\boldsymbol{ u}=\boldsymbol{ u}_0+\boldsymbol{ u}'$, $p=p_0+p'$, and $\boldsymbol{ B}=\boldsymbol{ B}_0+\boldsymbol{ B}'$ and leaving only the terms of first order with respect to the primed quantities:
\ba{L1}
&\partial_t\boldsymbol{ u}'+\boldsymbol{ u}_0 \bcdot \bnabla\boldsymbol{ u}'+\boldsymbol{ u}' \bcdot \bnabla\boldsymbol{ u}_0
-\frac{1}{\rho\mu_0}\left(\boldsymbol{ B}_0 \bcdot \bnabla\boldsymbol{ B}'{+}\boldsymbol{ B}' \bcdot \bnabla\boldsymbol{ B}_0 \right)-\nu \bnabla^2 \boldsymbol{ u}'=&\nn\\
&-\frac{1}{\rho}\bnabla p'-
\frac{1}{\rho \mu_0}\bnabla(\boldsymbol{ B}_0 \bcdot \boldsymbol{ B}'),&\nn\\
&\partial_t \boldsymbol{ B}'+\boldsymbol{ u}_0 \bcdot \bnabla\boldsymbol{ B}'+\boldsymbol{ u}' \bcdot \bnabla\boldsymbol{ B}_0-
\boldsymbol{ B}_0\bcdot \bnabla\boldsymbol{ u}'-\boldsymbol{ B}'\bcdot \bnabla\boldsymbol{ u}_0-\eta \bnabla^2 \boldsymbol{ B}'=0,&
\ea
where the perturbations fulfil the constraints
\be{L2}
\bnabla \bcdot \boldsymbol{ u}' = 0,\quad  \bnabla \bcdot \boldsymbol{ B}'=0.
\ee
Note that by adding/subtracting the second of equations \rf{L1} divided by $\sqrt{\rho \mu_0}$ to/from the first one yields the linearized equations in a more symmetrical Elsasser form; in the ideal MHD case they were derived, e.g. by \cite{FV95}.

Introducing the gradients of the background fields represented by the two $3 \times 3$ matrices
\ba{L3a}
&\mathcal{U}(R)=\bnabla \boldsymbol{ u}_0=\Omega\left(
                                \begin{array}{ccc}
                                  0 & -1 & 0 \\
                                  1+2{\rm Ro} & 0 & 0 \\
                                  0 & 0 & 0 \\
                                \end{array}
                              \right),& \nn\\
&\mathcal{B}(R)=\bnabla \boldsymbol{ B}_0=\frac{B_{\phi}^0}{R}\left(
 \begin{array}{ccc}
                                                                              0 & -1 & 0 \\
                                                                              1+2{\rm Rb} & 0 & 0 \\
                                                                              0 & 0 & 0 \\
                                                                            \end{array}
                                                                          \right),&
\ea
we write the linearized equations of motion in the form
\ba{L4}
&(\partial_t+\mathcal{U}+\boldsymbol{ u}_0 \bcdot \bnabla)\boldsymbol{ u}'
-\frac{1}{\rho\mu_0}\left(\mathcal{B}+\boldsymbol{ B}_0 \bcdot \bnabla\right)\boldsymbol{ B}' -\nu \bnabla^2 \boldsymbol{ u}'
+\frac{1}{\rho}\bnabla p'+
\frac{1}{\rho \mu_0}\bnabla(\boldsymbol{ B}_0 \bcdot \boldsymbol{ B}')=0,&\nn\\
&(\partial_t -\mathcal{U}+\boldsymbol{ u}_0 \bcdot \bnabla)\boldsymbol{ B}'+(\mathcal{B}-
\boldsymbol{B}_0\bcdot \bnabla)\boldsymbol{ u}'-\eta \bnabla^2 \boldsymbol{ B}'=0.&
\ea


\section{Geometrical optics equations}

Following \cite{E81}, \cite{L89,L91} and \cite{FL03}, we seek for solutions of the linearized equations \rf{L4} in terms of the formal asymptotic series with respect to the small parameter $\epsilon$, $0<\epsilon\ll 1$:
\ba{g1}
&\boldsymbol{ u}'(\boldsymbol{ x},t,\epsilon)=e^{i\Phi(\boldsymbol{ x},t)/\epsilon}\left(\boldsymbol{ u}^{(0)}(\boldsymbol{ x},t)+\epsilon\boldsymbol{ u}^{(1)}(\boldsymbol{ x},t) \right)+\epsilon \boldsymbol{u}^{(r)}(\boldsymbol{ x},t),&\nn \\
&\boldsymbol{ B}'(\boldsymbol{ x},t,\epsilon)=e^{i\Phi(\boldsymbol{ x},t)/\epsilon}\left(\boldsymbol{ B}^{(0)}(\boldsymbol{ x},t)+\epsilon\boldsymbol{ B}^{(1)}(\boldsymbol{ x},t) \right)+\epsilon \boldsymbol{ B}^{(r)}(\boldsymbol{ x},t),&\nn \\
&p'(\boldsymbol{ x},t,\epsilon)=e^{i\Phi(\boldsymbol{ x},t)/\epsilon}\left(p^{(0)}(\boldsymbol{ x},t)+\epsilon p^{(1)}(\boldsymbol{ x},t) \right)+\epsilon p^{(r)}(\boldsymbol{ x},t),&
\ea
where $\boldsymbol{ x}$ is a vector of coordinates, $\Phi$ is a real-valued scalar function that represents the `fast' phase of oscillations, and $\boldsymbol{ u}^{(j)}$, $\boldsymbol{ B}^{(j)}$, and $p^{(j)}$, $j=0,1,r$ are `slow' complex-valued amplitudes with the index $r$ denoting residual terms.

Following \cite{LS87}, \cite{L91}, \cite{DS92}, and \cite{EY95} we assume further in the text that $\nu=\epsilon^2\tilde \nu$ and $\eta=\epsilon^2\tilde \eta$ and introduce the derivative
along the fluid stream lines
\be{g2}
\frac{D}{Dt}:=\partial_t + \boldsymbol{ u}_0 \bcdot \bnabla.
\ee

Substituting expansions \rf{g1} into equations \rf{L4}, taking into account the identity
\be{g2a}
(\boldsymbol{ A}\bcdot \bnabla)\Phi \boldsymbol{ B}=(\boldsymbol{ A} \bcdot \bnabla \Phi)\boldsymbol{ B}+\Phi (\boldsymbol{ A}\bcdot \bnabla) \boldsymbol{ B},
\ee
as well as the relation
\be{A2}
\bnabla^2(\boldsymbol{  u}')=e^{i\frac{\Phi}{\epsilon}}\left(\bnabla^2+i\frac{2}{\epsilon}(\bnabla \Phi\bcdot \bnabla)+
i\frac{ \bnabla^2 \Phi}{\epsilon} -\frac{(\bnabla \Phi)^2}{\epsilon^2}\right) (\boldsymbol{  u}^{(0)} +\epsilon\boldsymbol{  u}^{(1)})+\epsilon\bnabla^2(\boldsymbol{  u}^{(r)}),
\ee
collecting terms at $\epsilon^{-1}$ and $\epsilon^0$ we arrive at the system of local partial differential equations (cf. \cite{FV95})
\ba{g3}
&\frac{D\Phi}{Dt}  \boldsymbol{ u}^{(0)} -
\frac{1}{\rho\mu_0}(\boldsymbol{ B}_0\bcdot \bnabla\Phi)\boldsymbol{ B}^{(0)}+\frac{\bnabla \Phi}{\rho}  \left(p^{(0)}+\frac{1}{\mu_0}(\boldsymbol{ B}_0\bcdot \boldsymbol{ B}^{(0)})\right)=
0,&\\
&\frac{D\Phi}{Dt} \boldsymbol{ B}^{(0)}-(\boldsymbol{ B}_0 \bcdot \bnabla \Phi) \boldsymbol{ u}^{(0)}=0,&\nn\\
&i\frac{D \Phi}{Dt}\boldsymbol{ u}^{(1)}+\left(\frac{D }{Dt}  + \tilde \nu(\bnabla \Phi)^2 + \mathcal{U}\right)\boldsymbol{ u}^{(0)}-\frac{1}{\rho \mu_0}\left(\mathcal{B}+\boldsymbol{ B}_0\bcdot \bnabla \right)\boldsymbol{ B}^{(0)}-\frac{i}{\rho \mu_0}(\boldsymbol{ B}_0 \bcdot \bnabla \Phi) \boldsymbol{ B}^{(1)}&\nn  \\
&+\frac{\bnabla}{\rho} \left(  p^{(0)}+\frac{1}{\mu_0}(\boldsymbol{ B}_0\bcdot \boldsymbol{ B}^{(0)})\right)+
i\frac{\bnabla \Phi}{\rho}\left(p^{(1)}+\frac{1}{\mu_0}(\boldsymbol{ B}_0\bcdot \boldsymbol{ B}^{(1)})\right)=0,&\nn\\
&i\frac{D\Phi}{Dt}\boldsymbol{ B}^{(1)}+\left(\frac{D}{Dt}+
\tilde \eta(\bnabla \Phi)^2 -\mathcal{U} \right) \boldsymbol{ B}^{(0)}+(\mathcal{B}-\boldsymbol{ B}_0\bcdot \bnabla)
\boldsymbol{ u}^{(0)}-i(\boldsymbol{ B}_0\bcdot \bnabla \Phi)\boldsymbol{ u}^{(1)}=0.&\nn
\ea

From the solenoidality conditions \rf{L2} it follows that
\ba{g4}
&\boldsymbol{ u}^{(0)}\bcdot \bnabla\Phi=0,\quad \bnabla\bcdot \boldsymbol{ u}^{(0)}+i\boldsymbol{ u}^{(1)}\bcdot \bnabla \Phi
=0,&\nn\\
&\boldsymbol{ B}^{(0)}\bcdot \bnabla\Phi=0,\quad \bnabla\bcdot \boldsymbol{ B}^{(0)}+i\boldsymbol{ B}^{(1)}\bcdot \bnabla \Phi
=0.&
\ea

Following \cite{L91} we take the dot product of the first two of the equations \rf{g3} with $\bnabla \Phi$, $\boldsymbol{ u}^{(0)}$, $\boldsymbol{ B}^{(0)}$ and,
in view of the constraints \rf{g4}, arrive at the system
\ba{g5}
&\frac{(\bnabla \Phi)^2}{\rho}\left(p^{(0)}  +\frac{1}{\mu_0}(\boldsymbol{ B}_0\bcdot \boldsymbol{ B}^{(0)})\right)=0,&\nn\\
&\frac{D\Phi}{Dt} \boldsymbol{ B}^{(0)}\bcdot \boldsymbol{ B}^{(0)}-(\boldsymbol{ B}_0 \bcdot \bnabla \Phi) \boldsymbol{ u}^{(0)}\bcdot \boldsymbol{ B}^{(0)}=0,&\nn\\
&\frac{D\Phi}{Dt} \boldsymbol{ B}^{(0)}\bcdot \boldsymbol{ u}^{(0)}-(\boldsymbol{ B}_0 \bcdot \bnabla \Phi) \boldsymbol{ u}^{(0)}\bcdot \boldsymbol{ u}^{(0)}=0,&\nn\\
&\frac{D\Phi}{Dt}  \boldsymbol{ u}^{(0)}\bcdot \boldsymbol{ B}^{(0)} -
\frac{1}{\rho\mu_0}\boldsymbol{ B}^{(0)}\bcdot \boldsymbol{ B}^{(0)}(\boldsymbol{ B}_0\bcdot \bnabla\Phi)=0,&\nn\\
&\frac{D\Phi}{Dt}  \boldsymbol{ u}^{(0)}\bcdot \boldsymbol{ u}^{(0)} -
\frac{1}{\rho\mu_0}\boldsymbol{ B}^{(0)}\bcdot \boldsymbol{ u}^{(0)}(\boldsymbol{ B}_0\bcdot \bnabla\Phi)=0,&
\ea
that has for $\bnabla \Phi \ne 0$, $\boldsymbol{ B}^{(0)}\ne 0$, and $\boldsymbol{ u}^{(0)}\ne 0$ a unique solution
\be{g6}
p^{(0)}  =-\frac{1}{\mu_0}(\boldsymbol{ B}_0\bcdot \boldsymbol{ B}^{(0)}),\quad \frac{D\Phi}{Dt}=0, \quad \boldsymbol{ B}_0\bcdot \bnabla\Phi=0.
\ee

With the use of the relations \rf{g6} we simplify the last two of the equations \rf{g3} as
\ba{g7}
&\left(\frac{D}{Dt}+\tilde \nu(\bnabla \Phi)^2 +\mathcal{U} \right) \boldsymbol{ u}^{(0)}
-
\frac{1}{\rho\mu_0}(\mathcal{B}+\boldsymbol{ B}_0\bcdot \bnabla)\boldsymbol{ B}^{(0)} =
-i\frac{\bnabla \Phi}{\rho} \left(p^{(1)}+\frac{1}{\mu_0}(\boldsymbol{ B}_0 \bcdot \boldsymbol{ B}^{(1)})\right),&\nn\\
&\left(\frac{D}{Dt}+\tilde \eta(\bnabla \Phi)^2 -\mathcal{U} \right) \boldsymbol{B}^{(0)}+(\mathcal{B}-\boldsymbol{ B}_0\bcdot \bnabla)
\boldsymbol{ u}^{(0)}=0.&
\ea
Multiplying the first of Eqs.~\rf{g7}  by $\bnabla \Phi$ from the left and then dividing both parts by $|\bnabla \Phi|^2$ we find an expression for $p^{(1)}+\frac{1}{\mu_0}(\boldsymbol{ B}_0 \bcdot \boldsymbol{ B}^{(1)})$. Substituting it back to the right hand side of the first of Eqs.~\rf{g7} and taking into account the constraints \rf{g4},  we eliminate the pressure terms and transform this equation into
\ba{g8}
&\left(\frac{D}{Dt}+\tilde \nu(\bnabla \Phi)^2 +\mathcal{U} \right) \boldsymbol{ u}^{(0)}
-
\frac{1}{\rho\mu_0}(\mathcal{B}+\boldsymbol{ B}_0\bcdot \bnabla)\boldsymbol{ B}^{(0)} &\nn\\
&=\frac{\bnabla \Phi}{|\bnabla \Phi|^2}\bcdot\left[\left(\frac{D}{Dt} +\mathcal{U} \right) \boldsymbol{ u}^{(0)}
-
\frac{1}{\rho\mu_0}(\mathcal{B}+\boldsymbol{ B}_0\bcdot \bnabla)\boldsymbol{ B}^{(0)} \right]{\bnabla \Phi},&
\ea
quite in accordance with the standard procedure described, e.g., in \cite{L91} and \cite{VF98}.
Differentiating the first of the identities \rf{g4} yields
\be{g9}
\frac{D}{Dt}( \bnabla \Phi\bcdot\boldsymbol{ u}^{(0)})=\frac{D \bnabla \Phi}{Dt}\bcdot\boldsymbol{ u}^{(0)}+\bnabla \Phi\bcdot\frac{D \boldsymbol{ u}^{(0)}}{Dt}=0.
\ee
Using the identity \rf{g9}, we re-write Eq.~\rf{g8} as follows
\ba{g10}
&\left(\frac{D}{Dt}+\tilde \nu(\bnabla \Phi)^2 +\mathcal{U} \right) \boldsymbol{ u}^{(0)}
-
\frac{1}{\rho\mu_0}(\mathcal{B}+\boldsymbol{ B}_0\bcdot \bnabla)\boldsymbol{ B}^{(0)} &\nn\\
&=\frac{\bnabla \Phi}{|\bnabla \Phi|^2}\bcdot\left[
\mathcal{U} \boldsymbol{ u}^{(0)}
-
\frac{1}{\rho\mu_0}(\mathcal{B}+\boldsymbol{ B}_0\bcdot \bnabla)\boldsymbol{ B}^{(0)} \right]\bnabla \Phi-\frac{\bnabla \Phi}{|\bnabla \Phi|^2}
\frac{D \bnabla \Phi}{Dt}\bcdot\boldsymbol{ u}^{(0)},\nn
\ea
Now we take the gradient of the identity $D \Phi / Dt=0$:
\ba{g11}
&\bnabla \partial_t \Phi+\bnabla (\boldsymbol{ u}_0 \bcdot \bnabla)\Phi=\partial_t \bnabla \Phi+(\boldsymbol{ u}_0 \bcdot \bnabla)\bnabla \Phi+\mathcal{U}^T\bnabla\Phi&\nn\\
&=\frac{D}{D t}\bnabla \Phi+\mathcal{U}^T\bnabla\Phi=0.&
\ea
Denoting $\boldsymbol{ k}=\bnabla \Phi$, we deduce from the phase equation \rf{g11} that
\be{g12}
\frac{D\boldsymbol{ k}}{D t}=-\mathcal{U}^T\boldsymbol{ k}.
\ee
Hence, the transport equations for  the amplitudes  \rf{g7} take the final form
\ba{g13}
&\frac{D\boldsymbol{ u}^{(0)}}{Dt}=-\left(\mathcal{I}-
2\frac{\boldsymbol{ k}\boldsymbol{ k}^T}{|\boldsymbol{ k}|^2} \right)\mathcal{U} \boldsymbol{ u}^{(0)}-\tilde \nu
|\boldsymbol{ k}|^2\boldsymbol{ u}^{(0)}
+
\frac{1}{\rho\mu_0}\left(\mathcal{I}-\frac{\boldsymbol{ k}\boldsymbol{ k}^T}{|\boldsymbol{ k}|^2}\right)
\left(\mathcal{B}+\boldsymbol{B}_0\bcdot \bnabla\right)\boldsymbol{ B}^{(0)},& \nn\\
&\frac{D \boldsymbol{ B}^{(0)}}{Dt}= \mathcal{U} \boldsymbol{ B}^{(0)}-\tilde \eta|\boldsymbol{ k}|^2\boldsymbol{ B}^{(0)}-
(\mathcal{B}-\boldsymbol{ B}_0\bcdot \bnabla) \boldsymbol{ u}^{(0)},&
\ea
where $\mathcal{I}$ is a $3\times 3$ identity matrix.

Eqs.~\rf{g13} are local partial differential equations analogous to those derived (in terms of Elsasser variables $\boldsymbol{ u}^{(0)}\pm {\boldsymbol{ B}^{(0)}}(\rho \mu_0)^{-1/2}$) by \cite{FV95}  who considered non-axisymmetric perturbation of a rotating flow of an ideal incompressible fluid which is a perfect electrical conductor in the presence of the azimuthal magnetic field of arbitrary radial and axial dependency, see Appendix~\ref{appD}.

In the absence of the magnetic field these equations can be treated as ordinary differential equations with respect to the convective derivative \rf{g2} and thus are reduced to that of \cite{EY95} who considered stability of the viscous TC-flow. Note that the same form of the transport equations (with the different matrix $\mathcal{U}$) appears in the studies of elliptical instability by \cite{LS87} and of three-dimensional local instabilities of more general viscous and inviscid basic flows by \cite{L91}, \cite{LH91}, and \cite{DS92}.

\section{Dispersion relation of AMRI}

In case when the axial field is absent, i.e. $B_z^0=0$, we can derive the dispersion relation of AMRI directly from the transport equations \rf{g13} according to the procedure described in \cite{FV95}.

Let the orthogonal unit vectors $\boldsymbol{e}_R(t)$, $\boldsymbol{e}_{\phi}(t)$, and $\boldsymbol{e}_z(t)$
form a basis in a cylindrical coordinate system moving along the fluid trajectory. With $\boldsymbol{k}(t)=k_R \boldsymbol{e}_R(t)+k_{\phi} \boldsymbol{ e}_{\phi}(t)+k_z \boldsymbol{ e}_z(t) $, $\boldsymbol{ u}(t)=u_R \boldsymbol{ e}_R(t)+u_{\phi} \boldsymbol{ e}_{\phi}(t)+u_z \boldsymbol{ e}_z(t) $, and with the matrix $\mathcal{U}$ from \rf{L3a}, we find
\be{ae1}
\dot{\boldsymbol{ e}}_R=\Omega(R){\bf e}_{\phi}, \quad \dot{\boldsymbol{ e}}_{\phi}=-\Omega(R)\boldsymbol{ e}_R.
\ee
Hence, the equation \rf{g12} in the coordinate form is
\be{ae2}
\dot k_R= -R \partial_R \Omega k_{\phi}, \quad \dot k_{\phi}=0, \quad \dot k_z=0.
\ee

According to \cite{EY95} and \cite{FV95}, in order to study physically relevant and potentially unstable modes we have to choose bounded and asymptotically non-decaying solutions of the system \rf{ae2}. These correspond to $k_{\phi}\equiv 0$ and $k_R$ and $k_z$ time-independent. Note that when $B_z^0=0$ this solution is compatible with the constraint $\boldsymbol{ B}_0\bcdot \boldsymbol{k}=0$ following from \rf{g6}.

Denoting $\alpha=k_z|\boldsymbol{ k}|^{-1}$, where $|\boldsymbol{ k} |^2=k_R^2+k_z^2$, we find that $k_R k_z^{-1}=\sqrt{1-\alpha^2}\alpha^{-1}$ we write the local partial differential
equations \rf{g13} for the amplitudes in the coordinate representation. Analogously to \cite{FV95}, we single out the equations for the radial and azimuthal components of the fluid velocity and magnetic field by using  the orthogonality condition ${B}_R^{(0)} k_R+{B}_z^{(0)}k_z=0$ that follows from \rf{g4}:
\ba{mri4a}
&\left(\partial_t {+} \Omega\partial_{\phi} {+}\widetilde{\nu} |\boldsymbol{ k} |^2\right) u_R^{(0)}   -2\alpha^2\Omega u_{\phi}^{(0)} +2\alpha^2\frac{B_{\phi}^0}{\rho \mu_0 R}B_{\phi}^{(0)}- \frac{B_{\phi}^0}{R} \frac{\partial_{\phi}B_R^{(0)}}{\rho \mu_0}=0,&\nn\\
&\left(\partial_t {+}\Omega\partial_{\phi}  +\widetilde{\nu} |\boldsymbol{ k} |^2\right) u_{\phi}^{(0)}+2\Omega(1{+} {\rm Ro})u_R^{(0)}-\frac{2}{\rho \mu_0}\frac{B_{\phi}^0}{R}(1{+}{\rm Rb})B_R^{(0)}- \frac{B_{\phi}^0}{R} \frac{\partial_{\phi} B_{\phi}^{(0)}}{\rho \mu_0}=0,&\nn\\
&\left(\partial_t {+}\Omega\partial_{\phi}  {+}\widetilde{\eta} |\boldsymbol{ k} |^2\right) B_R^{(0)}-\frac{B_{\phi}^0}{R} \partial_{\phi}u_R^{(0)}=0,&\nn\\
&\left(\partial_t {+}\Omega\partial_{\phi}   {+}\widetilde{\eta} |\boldsymbol{ k} |^2\right) B_{\phi}^{(0)}-2\Omega {\rm Ro}B_R^{(0)}+2{\rm Rb}\frac{B_{\phi}^0}{R}u_R^{(0)}-\frac{B_{\phi}^0}{R}\partial_{\phi}u_{\phi}^{(0)}=0.&
\ea

According to \cite{FV95} a natural step is to seek for a solution to Eqs.~\rf{mri4a} in the modal form $\boldsymbol{ u}^{(0)}=\widehat{\boldsymbol{ u}} e^{\gamma t +im\phi}$, $\boldsymbol{ B}^{(0)}=\widehat{\boldsymbol{ B}} e^{\gamma t +im\phi}$  in order to end up with the dispersion relation for the transport equations in case of AMRI:
\ba{mri4ab}
&(\gamma +  im \Omega  +\widetilde{\nu} |\boldsymbol{ k} |^2) \widehat{u}_R   -2\alpha^2\Omega \widehat{u}_{\phi} +2\alpha^2\frac{B_{\phi}^0}{\rho \mu_0 R}\widehat{B}_{\phi}-im \frac{B_{\phi}^0}{R}\frac{\widehat{B}_R}{\rho \mu_0}=0,&\nn\\
&(\gamma +im \Omega +\widetilde{\nu} |\boldsymbol{ k} |^2) \widehat{u}_{\phi}  +2\Omega(1+ {\rm Ro})\widehat{u}_R-\frac{2}{\rho \mu_0}\frac{B_{\phi}^0}{R}(1+{\rm Rb})\widehat{B}_R-im\frac{B_{\phi}^0}{R}\frac{\widehat{B}_{\phi}}{\rho \mu_0} =0,&\nn\\
&(\gamma + im \Omega  +\widetilde{\eta} |\boldsymbol{ k} |^2) \widehat{B}_R-im\frac{B_{\phi}^0}{R} \widehat{u}_R  =0,&\nn\\
&(\gamma +im \Omega  +\widetilde{\eta} |\boldsymbol{ k} |^2) \widehat{B}_{\phi} -2\Omega {\rm Ro}\widehat{B}_R+2{\rm Rb}\frac{B_{\phi}^0}{R}\widehat{u}_R-im \frac{B_{\phi}^0}{R}\widehat{u}_{\phi}  =0.&
\ea

Introducing the viscous and resistive frequencies and the $\rm Alfv\acute{e}n$ angular velocity \citep{OP1996} corresponding to the azimuthal magnetic field:
\be{freq}
\omega_{\nu}=\widetilde{\nu} |\boldsymbol{ k}|^2,\quad \omega_{\eta}=\widetilde{\eta}|\boldsymbol{ k}|^2,\quad \omega_{A_{\phi}}=\frac{B_{\phi}^0}{R\sqrt{\rho\mu_0}},
\ee
so that ${\rm Rb}$ is simply
\be{freqa}
{\rm Rb}=\frac{R}{2\omega_{A_{\phi}}} \partial_R \omega_{A_{\phi}},
\ee
we write the amplitude equations \rf{mri4ab} as ${\bf A}{\bf z}=\gamma{\bf z}$, where ${\bf z}=(\widehat{u}_R,\widehat{u}_{\phi},\widehat{B}_R,\widehat{B}_{\phi})^T$ and
\be{mri11a}
{\bf A}=\left(
    \begin{array}{cccc}
      -im\Omega-\omega_{\nu} & 2\alpha^2\Omega & i  \frac{m \omega_{A_{\phi}}}{\sqrt{\rho \mu_0} } & -\frac{2\omega_{A_{\phi}}\alpha^2}{\sqrt{\rho \mu_0} } \\
      -2\Omega(1+ {\rm Ro}) & -im\Omega-\omega_{\nu} & \frac{2\omega_{A_{\phi}}}{\sqrt{\rho \mu_0}}(1+{\rm Rb}) & i\frac{m \omega_{A_{\phi}}  }{\sqrt{\rho \mu_0}} \\
      i m \omega_{A_{\phi}}\sqrt{\rho \mu_0}  & 0 & -im\Omega-\omega_{\eta} & 0 \\
      -2{\omega_{A_{\phi}}}{\rm Rb}{\sqrt{\rho \mu_0} } & i m \omega_{A_{\phi}}\sqrt{\rho \mu_0} & 2\Omega {\rm Ro} & -im\Omega-\omega_{\eta} \\
    \end{array}
  \right).
\ee
In the case when $\omega_{\nu}=0$ and $\omega_{\eta}=0$  the coefficients of the characteristic polynomial of the matrix \rf{mri11a} exactly coincide with the dispersion relation of ideal AMRI  derived by \cite{FV95}, see Appendix~\ref{appD} for the detailed comparison.

\section{Dispersion relation of HMRI}

It is instructive to derive the dispersion relation of HMRI by another approach similar to that of \cite{KS10}.

We write the linearized equations \rf{L4} in cylindrical coordinates $(R,\phi,z)$ and assume that $\boldsymbol{u}'=\widetilde{\boldsymbol{u}}(R)e^{\gamma t+im\phi+ik_z z}$,
$\boldsymbol{B}'=\widetilde{\boldsymbol{B}}(R)e^{\gamma t+im\phi+ik_z z}$, and $p'=\widetilde p(R) e^{\gamma t+im\phi+ik_z z}$. This yields a set of equations containing only functions of $R$ and radial derivatives:
\ba{navs}
&\left(
  \begin{array}{ccc}
    \gamma+im\Omega-\nu D & -2\Omega & 0 \\
    2(1+{\rm Ro}) & \gamma+im\Omega-\nu D & 0 \\
    0 & 0 & \gamma+im\Omega-\nu D \\
  \end{array}
\right)\widetilde{\boldsymbol{u}}-\frac{\nu}{R^2}\left(
                                        \begin{array}{ccc}
                                          -1 & -2im & 0 \\
                                          2im & -1 & 0 \\
                                          0 & 0 & 0 \\
                                        \end{array}
                                      \right)\widetilde{\boldsymbol{u}}&\nn\\
&- \frac{B_{\phi}^0}{\rho \mu_0R}\left(
                       \begin{array}{ccc}
                         im & -2 & 0 \\
                         2(1+{\rm Rb}) & im & 0 \\
                         0 & 0 & im \\
                       \end{array}
                     \right)\widetilde{\boldsymbol{B}}-
\frac{B_z^0}{\rho \mu_0}\left(
       \begin{array}{ccc}
         ik_z & 0 & 0 \\
         0 & ik_z & 0 \\
         0 & 0 & ik_z \\
       \end{array}
     \right)\widetilde{\boldsymbol{B}}&\nn\\
&+\left(
     \begin{array}{ccc}
       \partial_R & 0 & 0 \\
       0 & \frac{im}{R} & 0 \\
       0 & 0 & ik_z \\
     \end{array}
   \right)\widetilde{p}+\frac{1}{\rho \mu_0}\left(
                                  \begin{array}{ccc}
                                    0 & \frac{B_{\phi}^0}{R}(1+2{\rm Rb})+B_{\phi}^0\partial_R & B_z^0 \partial_R \\
                                    0 & \frac{im}{R} B_{\phi}^0 & \frac{im}{R} B_z^0 \\
                                    0 & ik_z B_{\phi}^0 & ik_z B_z^0 \\
                                  \end{array}
                                \right)\widetilde{\boldsymbol{B}}=0,
&
\ea
\ba{indu}
&\left(
  \begin{array}{ccc}
    \gamma+im\Omega-\eta D & 0 & 0 \\
    -2\Omega{\rm Ro} & \gamma+im\Omega-\eta D & 0 \\
    0 & 0 & \gamma+im\Omega-\eta D \\
  \end{array}
\right)\widetilde{\boldsymbol{B}}-\frac{\eta}{R^2}\left(
                                        \begin{array}{ccc}
                                          -1 & -2im & 0 \\
                                          2im & -1 & 0 \\
                                          0 & 0 & 0 \\
                                        \end{array}
                                      \right)\widetilde{\boldsymbol{B}}&\nn\\
&-\frac{B_{\phi}^0}{R}\left(
                       \begin{array}{ccc}
                         im & 0 & 0 \\
                         -2{\rm Rb} & im & 0 \\
                         0 & 0 & im \\
                       \end{array}
                     \right)\widetilde{\boldsymbol{u}}-
B_z^0\left(
       \begin{array}{ccc}
         ik_z & 0 & 0 \\
         0 & ik_z & 0 \\
         0 & 0 & ik_z \\
       \end{array}
     \right)\widetilde{\boldsymbol{u}}=0,
     &
\ea
where
$
D:=\partial_R^2+\frac{1}{R}\partial_R-\frac{m^2}{R^2}-k_z^2
$. The same operation applied to \rf{L2} yields
\be{solk}
\frac{\widetilde{u}_R}{R}+\partial_R\widetilde{u}_R+\frac{im}{R}\widetilde{u}_{\phi}+ik_z\widetilde{u}_z=0,\quad
\frac{\widetilde{B}_R}{R}+\partial_R\widetilde{B}_R+\frac{im}{R}\widetilde{B}_{\phi}+ik_z\widetilde{B}_z=0.
\ee
Note that the third equation of \rf{indu} is already separated from the others.

Substituting $\widetilde{\boldsymbol{u}}(R)=\widehat{\boldsymbol{u}}e^{ik_R R}$ and $\widetilde{\boldsymbol{B}}(R)=\widehat{\boldsymbol{B}}e^{ik_R R}$ into \rf{indu} and assuming that $R$ is fixed for a local analysis, we introduce into the resulting equations the parameters
\be{freq}
  \omega_{A_z}=\frac{k_z {B_z^0}}{\sqrt{\rho \mu_0}},\quad \omega_{A_{\phi}}=\frac{B_{\phi}^0}{R\sqrt{\rho\mu_0}},\quad\omega_{\nu}=\nu |\boldsymbol{ k}|^2,\quad \omega_{\eta}=\eta|\boldsymbol{ k}|^2,\quad |\boldsymbol{ k}|^2=k_R^2+k_z^2,
\ee
and finally set in these re-parameterized equations $R=(\epsilon k_R)^{-1}$, where $\epsilon:=(Rk_R)^{-1}\ll 1$ is a small parameter. After expanding the result in the powers of $\epsilon$, we find that the first two of equations \rf{indu} take the form
\ba{fitoeq}
&-i\sqrt{\rho\mu_0}(m\omega_{A_{\phi}}+\omega_{A_z})\widehat u_R+(\gamma+\omega_{\eta}+im\Omega)\widehat B_R=O(\epsilon),&\\
&2\omega_{A_{\phi}}{\rm Rb}\sqrt{\rho\mu_0}\widehat u_R-i(m\omega_{A_{\phi}}+\omega_{A_z})\sqrt{\rho\mu_0} \widehat u_{\phi}-2\Omega{\rm Ro}\widehat B_R+(\gamma+\omega_{\eta}+im\Omega )\widehat B_{\phi}=O(\epsilon).&\nn
\ea

Next, we express the pressure term from the third of equations \rf{navs} as follows:
\be{press}
\widetilde{p} = \frac{i}{k_z}\left((\gamma +i m \Omega-\nu D) \widetilde{u}_z-\frac{im B_{\phi}^0 }{\rho\mu_0 R}\widetilde{B}_z+ \frac{i k_z B_{\phi}^0}{\rho\mu_0}\widetilde{B}_{\phi}\right).
\ee
Analogously, finding $\widetilde{u}_z$ and $\widetilde{B}_z$ from \rf{solk} we substitute these terms together with the pressure term \rf{press} into the first two of equations \rf{navs}, assume in the result that $\widetilde{\boldsymbol{u}}(R)=\widehat{\boldsymbol{u}}e^{ik_R R}$ and $\widetilde{\boldsymbol{B}}(R)=\widehat{\boldsymbol{B}}e^{ik_R R}$ and take into account expressions \rf{freq}. Setting then $R=(\epsilon k_R)^{-1}$ in the equations and expanding them in $\epsilon$, we obtain
\ba{setoeq}
&(\gamma+\omega_{\nu}+im\Omega)\frac{k_R^2+k_z^2}{k_z^2}\widehat{u}_R-2\Omega \widehat{u}_{\phi}-i\frac{m\omega_{A_{\phi}}+\omega_{A_z}}{\sqrt{\rho\mu_0}}\frac{k_R^2+k_z^2}{k_z^2}\widehat{B}_R+\frac{2\omega_{A_{\phi}}}{\sqrt{\rho\mu_0}}\widehat{B}_{\phi}
=O(\epsilon),&\\
&2\Omega(1+{\rm Ro})\widehat{u}_R+(\gamma+\omega_{\nu}+i m\Omega)\widehat{u}_{\phi}-2(1+{\rm Rb})\frac{\omega_{A_{\phi}}}{\sqrt{\rho\mu_0}}\widehat{B}_R-i\frac{m\omega_{Af}+\omega_{A_z}}{\sqrt{\rho\mu_0}}\widehat{B}_{\phi}=O(\epsilon).&\nn
\ea
The equations \rf{setoeq} and \rf{fitoeq} yield a system ${\bf H}{\bf z}=\gamma{\bf z}$ with ${\bf z}=(\widehat{u}_R,\widehat{u}_{\phi},\widehat{B}_R,\widehat{B}_{\phi})^T$ and (cf. \cite{KS13,SB2014})
\be{mri11}
{\bf H}=\left(
    \begin{array}{cccc}
      -im\Omega-\omega_{\nu} & 2\alpha^2\Omega & i  \frac{m \omega_{A_{\phi}}+{\omega_{A_z}}}{\sqrt{\rho \mu_0} } & -\frac{2\omega_{A_{\phi}}\alpha^2}{\sqrt{\rho \mu_0} } \\
      -2\Omega(1+ {\rm Ro}) & -im\Omega-\omega_{\nu} & \frac{2\omega_{A_{\phi}}}{\sqrt{\rho \mu_0}}(1+{\rm Rb}) & i\frac{m \omega_{A_{\phi}}+{\omega_{A_z}}  }{\sqrt{\rho \mu_0}} \\
      i(m \omega_{A_{\phi}}+{\omega_{A_z}})\sqrt{\rho \mu_0}  & 0 & -im\Omega-\omega_{\eta} & 0 \\
      -2{\omega_{A_{\phi}}}{\rm Rb}{\sqrt{\rho \mu_0} } & i(m \omega_{A_{\phi}}+{\omega_{A_z}})\sqrt{\rho \mu_0} & 2\Omega {\rm Ro} & -im\Omega-\omega_{\eta} \\
    \end{array}
  \right),
\ee
where $\alpha=k_z|\boldsymbol{ k}|^{-1}$ and $\omega_{A_z}$ is the $\rm Alfv\acute{e}n$ frequency associated with the axial component of the magnetic field. When $\omega_{A_z}=0$, the matrix \rf{mri11} coincides with the matrix \rf{mri11a}.

The dispersion relation of HMRI
\be{dr1}
p(\gamma):=\det({\bf H}-\gamma {\bf I})=0,
\ee
where $\bf I$ is the $4\times4$ identity matrix, is
the fourth-order polynomial \rf{dr1}
\be{dr2}
p(\gamma)=(a_0+ib_0)\gamma^4+(a_1+ib_1)\gamma^3+(a_2+ib_2)\gamma^2+(a_3+ib_3)\gamma+a_4+ib_4
\ee
with complex coefficients that are given in explicit form by equation \rf{cp2} in Appendix~\ref{appC}.

In the presence of the magnetic fields with ${\rm Rb}=-1$, the dispersion relation \rf{dr1} reduces to that derived by \cite{KS12b} and in the case when ${\rm Rb}=-1$ and $m = 0$ to that of \cite{KS10}. In the ideal MHD case this dispersion relation was obtained already  by \cite{OP1996}.
In the absence of the magnetic fields, the
dispersion relation \rf{dr1} reduces to
that derived in the narrow-gap approximation  by \cite{KGD1966} for the non-axisymmetric
perturbations of the hydrodynamic TC flow. Explicitly, this connection is established in Appendix~\ref{appB}.

\section{Connection to known stability and instability criteria}
Let us compose a matrix filled in with the coefficients of the complex polynomial \rf{dr2}
\be{bim}
{\bf B}=\left(
  \begin{array}{rrrrrrrr}
    a_4 & -b_4 & 0 & 0 & 0 & 0 & 0 & 0 \\
    b_3 & a_3 & a_4 & -b_4 & 0 & 0 & 0 & 0 \\
    -a_2 & b_2 & b_3 & a_3 & a_4 & -b_4 & 0 & 0 \\
    -b_1 & -a_1 & -a_2 & b_2 & b_3 & a_3 & a_4 & -b_4 \\
    a_0 & -b_0 & -b_1 & -a_1 & -a_2 & b_2 & b_3 & a_3 \\
    0 & 0 & a_0 & -b_0 & -b_1 & -a_1 & -a_2 & b_2 \\
    0 & 0 & 0 & 0 & a_0 & -b_0 & -b_1 & -a_1 \\
    0 & 0 & 0 & 0 & 0 & 0 & a_0 & -b_0 \\
  \end{array}
\right).
\ee
The necessary and sufficient condition for stability by \cite{Bilharz44} requires positiveness of the determinants of all the four diagonal sub-matrices of even order of
the matrix $\bf B$
\be{bim1}
m_1=\det \left(
      \begin{array}{cc}
        a_4 & -b_4 \\
        b_3 & a_3 \\
      \end{array}
    \right)>0,\quad \ldots,\quad m_4=\det {\bf B}>0
\ee
in order that all the roots of the complex polynomial \rf{dr2} have negative real parts \citep{K2013dg}.
Although generally the Bilharz stability condition for the fourth-order complex polynomial consists of four separate criteria \rf{bim1}, it has been verified both numerically and analytically via analysis of the growth rates in \citep{KS10,KS13} that it is the last one, $m_4>0$, that determines stability for all considered cases.

The dispersion relation \rf{dr1} and stability criteria \rf{bim1} may look complicated at the first glance. In order to familiarize the reader with it, we first demonstrate that they easily yield a number of classical results, see e.g. \citep{AH1973,A1978}.

In the ideal case with $\omega_{\nu}=0$, and $\omega_{\eta}=0$ the dispersion relation \rf{dr1} takes the form
\ba{full}
p(\gamma)&=&4\alpha^2(\omega_{A_{\phi}}^2{\rm Rb}-\Omega^2{\rm Ro})\left((i\gamma-m\Omega)^2-(m\omega_{A_{\phi}}+\omega_{A_z})^2\right)\nn\\
&-&4\alpha^2\left(\Omega(i\gamma-m\Omega)+\omega_{A_{\phi}}(m\omega_{A_{\phi}}+\omega_{A_z})\right)^2\nn
+\left((i\gamma-m\Omega)^2-(m\omega_{A_{\phi}}+\omega_{A_z})^2\right)^2=0.\nn
\ea
At $\alpha=1$ this is exactly the dispersion relation by \cite{OP1996}. At $\omega_{A_{\phi}}=0$ and $m=0$ this is
the dispersion relation of SMRI by \cite{BH1992}. At $\omega_{\nu}=0$, $\omega_{\eta}=0$, $\omega_{A_z}=0$, and $\omega_{A_{\phi}}=0$
the roots of the dispersion relation \rf{dr1} are
$$
\gamma_{1,2}=-i m \Omega \pm i 2\alpha \Omega\sqrt{1+{\rm Ro}},\quad \gamma_{3,4}=- i m \Omega.
$$
At $\omega_{\nu}=0$, $\omega_{\eta}=0$, and $\omega_{A_{\phi}}^2{\rm Rb}=\Omega^2{\rm Ro}$ the roots are pure imaginary
\ba{vech}
&\gamma_{1,2}=-i(m+\alpha)\Omega\pm i\sqrt{(m+\alpha)^2\omega^2_{A_{\phi}}+\alpha^2(\Omega^2-\omega^2_{A_{\phi}})},&\nn\\
&\gamma_{3,4}=-i(m-\alpha)\Omega\pm i\sqrt{(m-\alpha)^2\omega^2_{A_{\phi}}+\alpha^2(\Omega^2-\omega^2_{A_{\phi}})}.&\nn
\ea

In the dissipative case, the threshold for the Standard MRI is given by the equation $m_4=0$ with $\omega_{A_{\phi}}=0$. This yields the expression found in \cite{KS12}
\be{bim2}
{\rm Ro}=-\frac{(\omega_{A_z}^2+\omega_{\nu}\omega_{\eta})^2+4\alpha^2\Omega^2\omega_{\eta}^2}{4\alpha^2\Omega^2(\omega_{A_z}^2+\omega_{\eta}^2)}.
\ee

When ${\rm Ro}=0$, $\Omega=0$, and $m=0$ we get from the equation $m_4=0$  an instability condition that extends that of \cite{CH61} (see also \cite{R10}) to the dissipative case
\be{bim3}
{\rm Rb}>\frac{(\omega_{\nu}\omega_{\eta}+\omega_{A_z}^2)^2-4\alpha^2\omega_{A_z}^2\omega_{A_{\phi}}^2}{4\alpha^2\omega_{A_{\phi}}^2(\omega_{\nu}\omega_{\eta}+\omega_{A_z}^2)}.
\ee

On the other hand, under the assumption that $\Omega=0$ and $\omega_{A_z}=0$ the dispersion relation \rf{dr1} is a real polynomial. Hence, by setting its constant term to zero we determine the condition for a static instability
\be{tay1om}
{\rm Rb}>\frac{(\omega_{\nu}\omega_{\eta}+m^2\omega_{A_{\phi}}^2)^2-4\alpha^2m^2\omega_{A_{\phi}}^4}{4\alpha^2\omega_{A_{\phi}}^2(\omega_{\nu}\omega_{\eta}+m^2\omega_{A_{\phi}}^2)}.
\ee
In the ideal case \rf{tay1om} reduces to the criterion of \cite{OP1996}: ${\rm Rb}>-1+\frac{m^2}{4\alpha^2}$.
At $\rm Rb=0$ the criterion \rf{tay1om} yields the standard \cite{TAYLER} instability at
\be{tay2om}
\omega_{A_{\phi}}>\sqrt{\frac{\omega_{\nu}\omega_{\eta}}{\alpha^2-(|m|- \alpha)^2}}.
\ee

Another particular case $\omega_{A_z}=0$ and $m=0$ yields the
following extension of the stability condition by \cite{Michael1954}:
\be{bim3}
{\rm Ro}>-1+{\rm Rb}\frac{\omega_{A_{\phi}}^2}{\Omega^2}\frac{\omega_{\nu}}{\omega_{\eta}}-\frac{\omega_{\nu}^2}{4\alpha^2\Omega^2}.
\ee
Choosing, additionally, $\omega_{A_{\phi}} = 0$, we reproduce the result of
\cite{EY95}
$$
{\rm Ro}>-1-\frac{\omega_{\nu}^2}{4  \alpha^2\Omega^2}.
$$

The ideal Michael criterion for stability of rotational flows in the presence of azimuthal magnetic fields with respect to axisymmetric perturbations
\be{bim4}
\frac{d}{d R}(\Omega^2 R^4)-\frac{R^4}{\rho \mu_0}\frac{d}{d R}\left( \frac{B_{\phi}^0}{R}\right)^2>0
\ee
or
\be{bim5}
{\rm Ro}>-1+{\rm Rb}\frac{\omega_{A_{\phi}}^2}{\Omega^2}
\ee
follows from Eq.~\rf{bim3} when $\omega_{\eta}=\omega_{\nu}$ and $\omega_{\nu} \rightarrow 0$
(see also \cite{HG1962}).

Finally, letting in the equation \rf{mri11}
\be{ces}
{\rm Rb} ={\rm Ro},\quad \Omega=\omega_{A_{\phi}}
\ee
and assuming $\omega_{\nu}=0$ and $\omega_{\eta}=0$, we find
that the dispersion relation \rf{dr1} has the roots
$$
\gamma_{1,2}=0,\quad \gamma_{3,4}=-2i\omega_{A_{\phi}}(m\pm \alpha),
$$
indicating marginal stability. Indeed, for the inviscid fluid of infinite electrical
conductivity with $P=const.$,  conditions \rf{ces}
define at ${\rm Rb} ={\rm Ro}=-1$  the \textit{Chandrasekhar equipartition
solution}--- a special solution with the $\rm Alfv\acute{e}n$ angular velocity identical to the fluid angular velocity---which is marginally stable \citep{Chandra56,CH61}.

\section{Dispersion relation in dimensionless parameters}

The dispersion relation \rf{dr1} with the matrix \rf{mri11} has the same roots as the equation
\be{drdim}
\det({\bf H}{\bf T}-\gamma {\bf T})=0,
\ee
where ${\bf T}=\diag(1,1,(\rho \mu_0)^{-1/2},(\rho \mu_0)^{-1/2})$.
Let us change in the equation \rf{drdim} the spectral parameter as $\gamma=\widetilde{\gamma} \sqrt{\omega_{\nu}\omega_{\eta}}$
and in addition to the hydrodynamic $({\rm Ro})$ and magnetic $({\rm Rb})$ Rossby numbers introduce the magnetic Prandtl number $({\rm Pm})$, the ratio of the Alfv\'en frequencies $(\beta)$, Reynolds $({\rm Re})$ and Hartmann $({\rm Ha})$ numbers as well as the modified azimuthal wavenumber $n$ as follows
\be{d3}
{\rm Pm}=\frac{\omega_{\nu}}{\omega_{\eta}},\quad
\beta=\alpha \frac{\omega_{A_{\phi}}}{\omega_{A_z}},\quad
{\rm Re}=\alpha\frac{\Omega}{\omega_{\nu}},\quad
{{\rm Ha}}=\frac{\omega_{A_z}}{\sqrt{\omega_{\nu}\omega_{\eta}}},\quad
n=\frac{m}{\alpha}.
\ee
Then, the dispersion relation \rf{drdim} transforms into
\be{d4}
p(\widetilde{\gamma})=\det\left(\widetilde {\bf H}-\frac{\widetilde{\gamma}}{\sqrt{\rm Pm}} {\bf I}\right)=0,
\ee
with
\be{d5}
\widetilde {\bf H}=\left(
  \begin{array}{cccc}
    -i n {\rm Re} -1 & 2\alpha {\rm Re}  & \frac{i {\rm Ha}(1+n\beta)}{\sqrt{\rm Pm}} & -\frac{2\alpha\beta{\rm Ha}}{\sqrt{\rm Pm}} \\
    -\frac{2{\rm Re}(1+{\rm Ro})}{\alpha} & -i n {\rm Re} -1 & \frac{2\beta{\rm Ha}(1+{\rm Rb})}{\alpha\sqrt{{\rm Pm}}} & \frac{i {\rm Ha}(1+n\beta)}{\sqrt{{\rm Pm}}} \\
    \frac{i {\rm Ha}(1+n\beta)}{\sqrt{\rm Pm}} & 0 & -i n {\rm Re} -\frac{1}{\rm Pm} & 0 \\
    \frac{-2\beta {\rm Ha}{\rm Rb}}{\alpha \sqrt{\rm Pm}} & \frac{i {\rm Ha}(1+n\beta)}{\sqrt{\rm Pm}} & \frac{2{\rm Re}{\rm Ro}}{\alpha} & -i n {\rm Re} -\frac{1}{\rm Pm}\\
  \end{array}
\right).
\ee
The coefficients of the polynomial \rf{d4} are explicitly given by Eq.~\rf{co1} in Appendix~\ref{appC}.

Next, we divide the equation \rf{d4}  by $\rm Re$ and introduce the eigenvalue parameter
\be{d5la}
\lambda=\frac{\widetilde \gamma}{{\rm Re}\sqrt{\rm Pm}}=\frac{\widetilde \gamma\sqrt{\omega_{\nu}\omega_{\eta}}}{\alpha \Omega}=\frac{\gamma}{\alpha\Omega}.
\ee
This results in the dispersion relation
\be{d5lb}
p(\lambda)=\det({\bf M}-\lambda{\bf I})=0,
\ee
with \citep{KS13}
\be{d5lc}
{\bf M}=\left(
  \begin{array}{cccc}
    -i n  -\frac{1}{\rm Re} & 2\alpha   & i (1{+}n\beta)\sqrt{\frac{\rm N}{\rm Rm}} & -2\alpha\beta\sqrt{\frac{\rm N}{\rm Rm}} \\
    -\frac{2(1+{\rm Ro})}{\alpha} & -i n  -\frac{1}{\rm Re} & \frac{2\beta(1+{\rm Rb})}{\alpha}\sqrt{\frac{\rm N}{\rm Rm}} & i(1{+}n\beta)\sqrt{\frac{\rm N}{\rm Rm}} \\
    i (1{+}n\beta)\sqrt{\frac{\rm N}{\rm Rm}} & 0 & -i n  -\frac{1}{\rm Rm} & 0 \\
    \frac{-2\beta {\rm Rb}}{\alpha}\sqrt{\frac{\rm N}{\rm Rm}} & i (1{+}n\beta)\sqrt{\frac{\rm N}{\rm Rm}} & \frac{2{\rm Ro}}{\alpha} & -i n  -\frac{1}{\rm Rm}\\
  \end{array}
\right).
\ee
Here, ${\rm N}={\rm Ha}^2/{\rm Re}$ is the Elsasser number (interaction parameter) and $\rm Rm=\rm Re Pm$ is the magnetic Reynolds number.
Explicit coefficients of the dispersion relation \rf{d5lb} can be found in equation \rf{co2} of the Appendix~\ref{appC}.
In the following, we will use the dispersion relations \rf{dr1}, \rf{d4}, and \rf{d5lb} with different parameterizations in order to facilitate physical interpretation and comparison with the results obtained in astrophysical, MHD, plasma physics, and hydrodynamical communities.


\section{Inductionless approximation}

In this section, we focus on the inductionless approximation by
setting the magnetic Prandtl number to zero \citep{Hollerbach2005,P11}
which is a reasonable approximation for liquid metal experiments
as well as for some colder parts of accretion disks.

\subsection{The threshold of instability}
\label{seq7_1}
Consider the dispersion relation \rf{d4}. Substitute its coefficients \rf{co1} to the matrix $\bf B$ in \rf{bim} and then calculate the determinants $m_i$ according to \rf{bim1}. Set ${\rm Pm}=0$ in the resulting expressions for $m_i$. For the coefficient $m_4=\det{\bf B}$ this leads to a great simplification and yields the instability threshold in a compact and closed form:
\be{fullset}
\frac{{\rm Ha}^4\beta^2(n\beta+1)^2{\rm Ro}^2-\left(C^2-4{\rm Ha}^4\beta^2(n\beta+1)^2\right)({\rm Ro}+1)}{(1+{\rm Ha}^2(n\beta+1)^2)^2-4{\rm Ha}^2\beta^2(1+{\rm Ha}^2(n\beta+1)^2){\rm Rb}-4{\rm Ha}^4\beta^2(n\beta+1)^2} =\frac{C^2}{4{\rm Re}^2},
\ee
where $C=1+{\rm Ha}^2((n\beta+1)^2-2\beta^2{\rm Rb})$. Further in the text we will derive this threshold also from the explicit expressions for the growth rate of the perturbation.

In the limit ${\rm Re} \rightarrow \infty$ the equation \rf{fullset} reduces to quadratic equation with respect to $\rm Ro$ in the numerator of its left side. Dividing this equation by ${\rm Ha}^4$ and then letting ${\rm Ha}\rightarrow \infty$ yields
\be{afterRe}
\beta^2(n\beta+1)^2{\rm Ro}^2-(((n\beta+1)^2-2\beta^2{\rm Rb})^2-4\beta^2(n \beta+1)^2)({\rm Ro}+1)=0
\ee
with the roots
\ba{ro1}
{\rm Ro}^{\pm}&=&-2+ \frac{(n\beta+1)^2-2\beta^2{\rm Rb}\pm\sqrt{((n\beta+1)^2-2\beta^2{\rm Rb})^2-4\beta^2(n\beta+1)^2}}{2\beta^2(n\beta+1)^2((n\beta+1)^2-2\beta^2{\rm Rb})^{-1}}.
\ea
At $n=0$ and ${\rm Rb=-1}$ the expression \rf{ro1} reduces to that of \cite{KS11}
\be{indm1}
{\rm Ro}^{\pm}=\frac{4\beta^4+1\pm(2\beta^2+1)\sqrt{4\beta^4+1}}
{2\beta^2}.
\ee
From \rf{indm1} one can find $\beta$ as a function of $\rm Ro$ to recover the results of \cite{P11}.

            \begin{figure}
    \begin{center}
    \includegraphics[angle=0, width=0.85\textwidth]{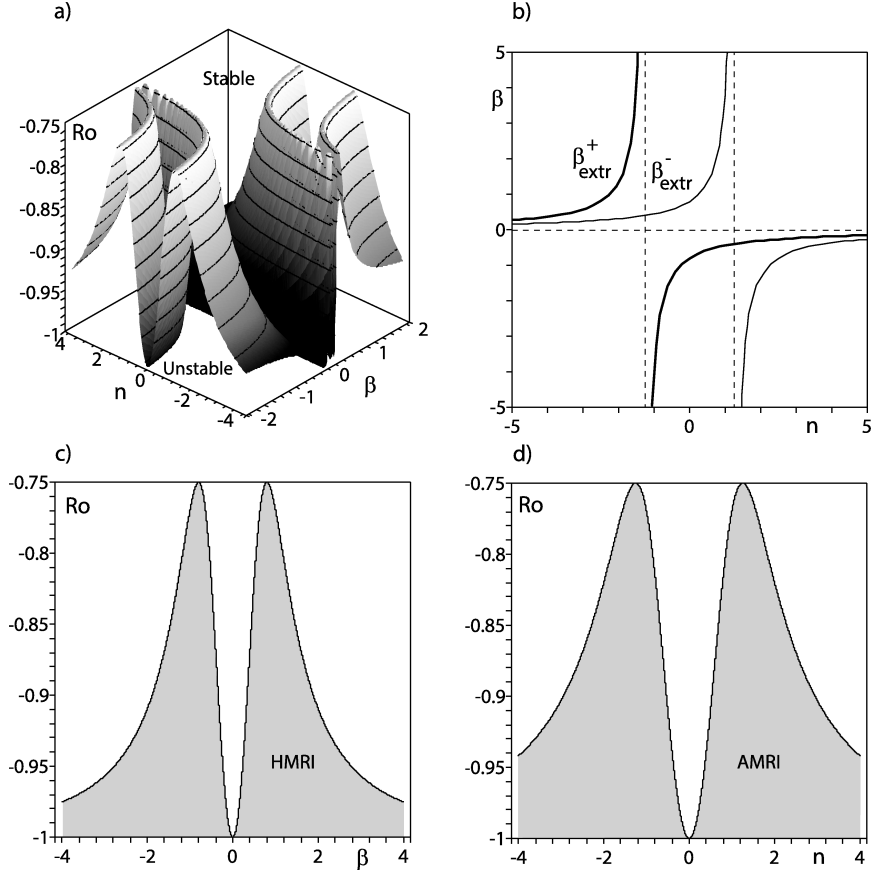}
    \end{center}
    \caption{(a) The function ${\rm Ro}^-(\beta,n)$ given by Eq.~\rf{ro1} at ${\rm Rb=-25/32}$. (b) Its extremizers \rf{liun} along which the critical hydrodynamic Rossby number attains the maximal value ${\rm Ro}^-=-3/4$ . The cross-section of the function at (c) $n=0$ (HMRI) and (d) at $\beta \rightarrow + \infty$ (AMRI). }
    \label{fig1}
    \end{figure}

\subsection{Extremal properties of the critical hydrodynamic Rossby number}
For a given magnetic Rossby number, $\rm Rb$, the functions ${\rm Ro}^{\pm}(n,\beta)$ take their extrema
\be{ro2}
{\rm Ro}_{\rm extr}^{\pm}=-2-4{\rm Rb}\pm2\sqrt{2{\rm Rb}(2{\rm Rb}+1)}
\ee
at the following extremal curves in the $(n,\beta)$-plane (cf. \cite{KS12b})
\be{liun}
\beta_{\rm extr}^+=\frac{-1}{n+\sqrt{-2{\rm Rb}}},\quad \beta_{\rm extr}^-=\frac{-1}{n-\sqrt{-2{\rm Rb}}},
\ee
see  Fig.~\ref{fig1}(b).
The branch ${\rm Ro}^-(\beta,n)$ is shown in Fig.~\ref{fig1}(a) for ${\rm Rb}=-25/32$ (this particular value has been chosen since it is
the minimum value which leads to destabilization of Keplerian profiles,
as we will see below).
The maximal value of the hydrodynamic Rossby number
is constant along the curves \rf{liun}, see Fig.~\ref{fig1}(a,b).

The cross-section of the function  ${\rm Ro}^-(\beta,n)$ at $n=0$ corresponding to the helical magnetorotational instability (HMRI) is plotted in Fig.~\ref{fig1}(c). In contrast,  Fig.~\ref{fig1}(d) shows the limit $\beta\rightarrow +\infty$
\be{infb}
{\rm Ro}^-(n,{\rm Rb}) = -2+(n^2-2{\rm Rb})\frac{n^2-2{\rm Rb}-\sqrt{(n^2-2{\rm Rb})^2-4n^2}}{2n^2}
\ee
corresponding to the azimuthal magnetorotational instability (AMRI).

The function \rf{infb} attains its maximal value ${\rm Ro}_{\rm extr}^{-}$ given by equation \rf{ro2} at
\be{indm3}
n=\pm \sqrt{-2{\rm Rb}}.
\ee
Since by definition \rf{d3} $n=\frac{m}{\alpha}$ where $\frac{k_z}{\sqrt{k_R^2+k_z^2}}=:\alpha \in [0,1]$,
then for ${\rm Rb}\in[-1, -0.5]$ the condition \rf{indm3} yields the only possible integer values of the azimuthal wavenumber:
$$
m=\pm 1.
$$
This means that the effect of the purely azimuthal magnetic field is most pronounced at the maximal possible range of variation of the hydrodynamic Rossby number for the lowest azimuthal modes with $m=\pm 1$ \citep{HTR10,KSF2014}.

\subsection{A band structure periodic in $\alpha$}
\begin{figure}
    \begin{center}
    \includegraphics[angle=0, width=0.95\textwidth]{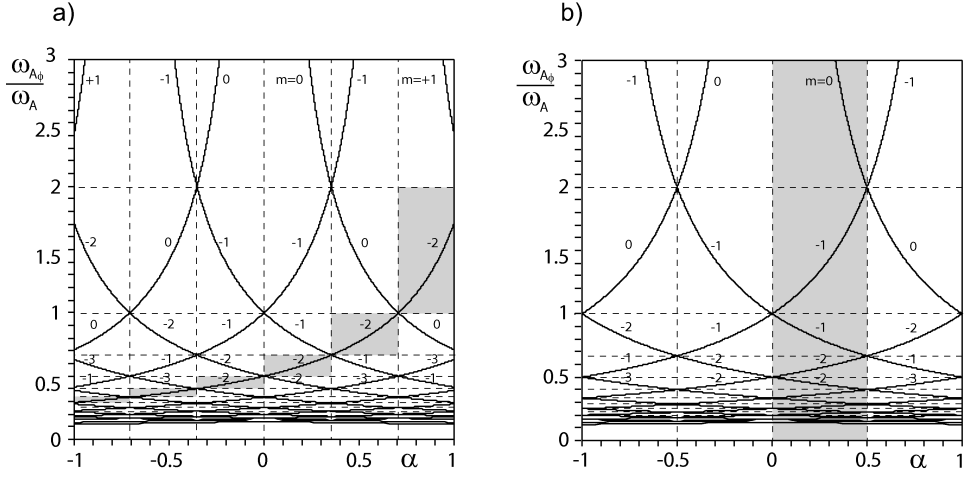}
    \end{center}
    \caption{The periodic in $\alpha$ band structure at (a) ${\rm Rb}=-1$ (with gray cells corresponding to $m=-2$) and (b) ${\rm Rb}=-0.5$ (with the gray stripe marking the first band).}
    \label{fig2}
    \end{figure}

What is the general dependence of the mode number $m$ on $\alpha$, $\rm Rb$, and the ratio $\omega_{A_{\phi}}/\omega_{A_z}$ of the azimuthal to the axial magnetic fields?
Let us express $\alpha$ from the  Eq.~\rf{liun} as
\be{alpha}
\alpha=\pm \left(m+\frac{\omega_{A_z}}{\omega_{A_{\phi}}}\right)\frac{1}{\sqrt{-2{\rm Rb}}}.
\ee
The ratio $\omega_{A_{\phi}}/\omega_{A_z}$ being plotted against $\alpha$ at different $m$ and $\rm Rb$ uncovers a regular pattern shown in Fig.~\ref{fig2}. The pattern is periodic in $\alpha$ with the period $1/\sqrt{-2{\rm Rb}}$. On the other hand, the dependence of $\alpha$ on the ratio of the magnetic fields is a continuous piecewise smooth function  inside every vertical `band' with the width $1/(2\sqrt{-2{\rm Rb}})$ in $\alpha$, see Fig.~\ref{fig2}(b). The vertical bands in Fig.~\ref{fig2} are further separated into the cells with the boundaries at
\be{cells}
\frac{\omega_{A_{\phi}}}{\omega_{A_z}}=\frac{2}{1},\frac{2}{2},\frac{2}{3},\frac{2}{4},\frac{2}{5},\frac{2}{6},\ldots .
\ee
Every particular cell in Fig.~\ref{fig2} corresponds to a unique integer azimuthal wavenumber $m$.
Note however that the cells with the same $m$ are grouped along the same hyperbolic curve given by equation \rf{alpha}.

The curves \rf{alpha} with the slopes of the same sign do not cross.
The crossings of two hyperbolic curves \rf{alpha} with the integer indices $m_1$ and $m_2$ and slopes of different sign happen at
\be{crossing}
\alpha=\frac{m_1-m_2}{2\sqrt{-2{\rm Rb}}},\quad {\rm and} \quad \frac{\omega_{A_{\phi}}}{\omega_{A_z}}=\frac{-2}{m_1+m_2},
\ee
which explains the sequence \rf{cells}.
Within one and the same band with $\alpha\ge 0$ the cells are encoded by the two interlacing series of integers, e.g. for the third band in Fig.~\ref{fig2}(a)
\be{band3}
m=+1,-2,0,-3,-1,-4,-2,-5,-3,-6,\ldots ,
\ee
which is a mixture of the sequences
$$
+1,0,-1,-2,-3,\ldots \quad {\rm and} \quad -2,-3,-4,-5,-6,\ldots ,
$$
whereas for the second band with $\alpha\ge 0$ in Fig.~\ref{fig2}(a) we have
\be{band2}
m=-1,0,-2,-1,-3,-2,-4,-3,-5,-4,\ldots ,
\ee
which corresponds to the two interlacing sequences
\be{band2a}
-1,-2,-3,-4,-5,\ldots \quad {\rm and} \quad 0,-1,-2,-3,-4,\ldots .
\ee

Therefore, at the crossings \rf{crossing} with $\alpha\ne 0$ the quadruplet of cells consists of two pairs; each pair corresponding to a different value of the azimuthal wavenumber $m$. At the crossings with $\alpha=0$ all the four cells in the quadruplet have the same $m$.

    \begin{figure}
    \begin{center}
    \includegraphics[angle=0, width=0.9\textwidth]{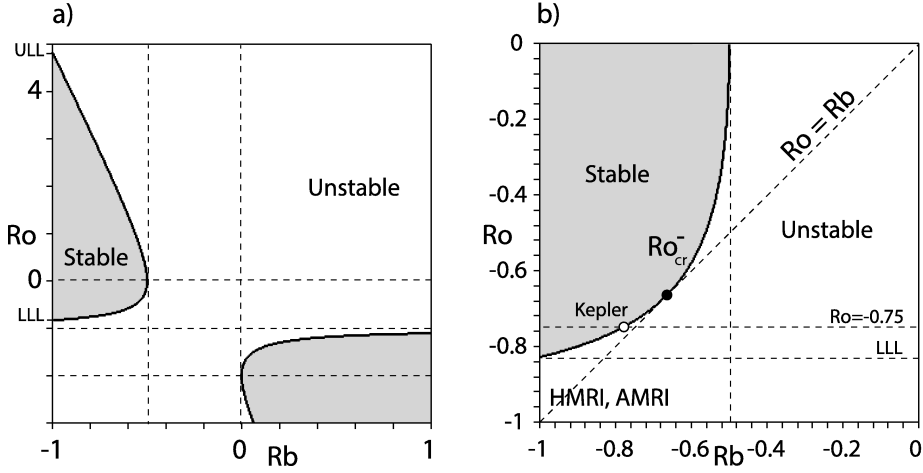}
    \end{center}
    \caption{(a) The lower (LLL) and the upper (ULL) Liu limits existing at ${\rm Rb}=-1$ are points on the quasi-hyperbolic curve \rf{hyp1} in the ${\rm Ro}-{\rm Rb}$ plane. (b) A scaled fragment of the limiting curve \rf{ro4} demonstrating that the inductionless forms of the magnetorotational instability exist above the lower Liu limit ${\rm Ro}_{\rm extr}^-=2-2\sqrt{2}$ when ${\rm Rb} > -25/32$. The open circle marks the Keplerian value for the Rossby number ${\rm Ro}=-3/4$ at ${\rm Rb} = -25/32$ whereas the black circle corresponds to ${\rm Ro}={\rm Rb}=-2/3$. The dashed diagonal is given by the equation ${\rm Ro}={\rm Rb}$.}
    \label{fig3}
    \end{figure}

\subsection{Continuation of the Liu limits to arbitrary $\rm Rb$}

The extrema \rf{ro2} can be represented in the form
\be{enf}
{\rm Ro}_{\rm extr}^{\pm}=-1-(\sqrt{2{\rm Rb}}\mp\sqrt{1+2{\rm Rb}})^2,
\ee
which is equivalent to the expression
\be{ro4}
{\rm Rb}=-\frac{1}{8}\frac{({\rm Ro}+2)^2}{{\rm Ro}+1}
\ee
that is universal in the sense that it contains only logarithmic derivatives of the radial profiles of the flow and the azimuthal field.

The particular case of equation \rf{ro4} at ${\rm Rb}= -1$ yields the result of \cite{Liu2006}, reproduced also
by \cite{KS11} and \cite{P11}.
Solving \rf{ro4} at ${\rm Rb}= -1$, we find that the critical Rossby numbers ${\rm Ro}({\rm Ha}, {\rm Re}, n, \beta)$ given by the equation \rf{fullset} and thus the instability domains lie at ${\rm Pm}=0$ and ${\rm Rb}= -1$ outside the stratum
$$
2-2\sqrt{2}=:{\rm Ro}_{\rm LLL}<{\rm Ro}({\rm Ha}, {\rm Re}, n, \beta)<{\rm Ro}_{\rm ULL}:=2+2\sqrt{2},
$$
where ${\rm Ro}_{\rm LLL}$ is the value of ${\rm Ro}^{-}_{\rm extr}$ at the lower Liu limit (LLL) and ${\rm Ro}_{\rm ULL}$ is the value of
${\rm Ro}^{+}_{\rm extr}$ at the upper Liu limit (ULL) corresponding to the critical values of $\beta$ given by the equation \rf{liun}.

Fig.~\ref{fig3}(a) shows how the LLL and ULL continue to the values of ${\rm Rb}\ne -1$. In fact, the LLL and the ULL are points at the
quasi-hyperbolic curve
\be{hyp1}
({\rm Rb}-{\rm Ro}-1)^2-({\rm Rb}+{\rm Ro}+1)^2=\frac{1}{2}({\rm Ro}+2)^2,
\ee
which is another representation of equation \rf{ro4}. At ${\rm Rb}=-1/2$ the branches ${\rm Ro}_{\rm extr}^-({\rm Rb})$ and ${\rm Ro}_{\rm extr}^+({\rm Rb})$ meet each other. Therefore, the
inductionless magnetorotational instability at negative $\rm Rb$ exists also at positive $\rm Ro$ when ${\rm Ro}>{\rm Ro}_{\rm extr}^+({\rm Rb})$, Fig.~\ref{fig3}(a). Notice also the second stability domain at ${\rm Rb}>0$ and ${\rm Ro}<-1$.

We see that in the inductionless case ${\rm Pm}=0$ when the Reynolds and Hartmann numbers subsequently tend to infinity as described in Section~\ref{seq7_1}
and $\beta$ and $n$ are under the constraints \rf{liun}, the maximal possible critical Rossby number ${\rm Ro}_{\rm extr}^-$ increases with the increase of ${\rm Rb}$. At
\be{ro5}
{\rm Rb}\ge -\frac{25}{32}=-0.78125
\ee
${\rm Ro}_{\rm extr}^-({\rm Rb})$  exceeds the critical value for the Keplerian flow: ${\rm Ro}_{\rm extr}^-\ge -3/4$.

Therefore, the possibility for $B_{\phi}(R)$ to depart from the profile $B_{\phi}(R)\propto R^{-1}$  allows us to break the conventional lower Liu limit and extend the inductionless versions of MRI to the velocity profiles $\Omega(R)$ as flat as the Keplerian one and even to less steep profiles, including that of the solid body rotation, i.e. ${\rm Ro=0}$, for ${\rm Rb}\ge-1/2$,  Fig.~\ref{fig3}(b).

            \begin{figure}
    \begin{center}
    \includegraphics[angle=0, width=0.85\textwidth]{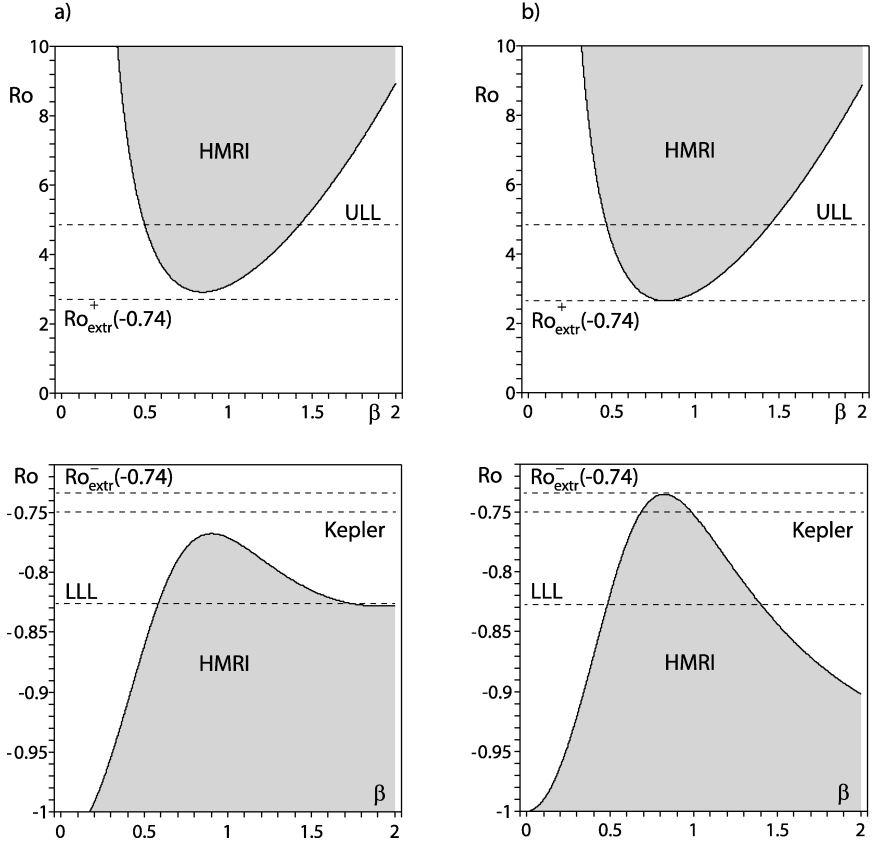}
    \end{center}
    \caption{Domains of the inductionless HMRI for $\rm Pm=0$, $n=0$, and $\rm Rb=-0.74$ when the Hartmann and Reynolds numbers follow the scaling law \rf{ro6} with $\beta$ satisfying the restraints \rf{liun}: (a) ${\rm Ha}=5$ and ${\rm Re}\approx92$; (b) ${\rm Ha}=100$ and ${\rm Re}\approx736158$. With the increase in $\rm Ha$ and $\rm Re$ the HMRI domain easily breaks the conventional Liu limits as well as the Kepler line $\rm Ro=-0.75$ and its boundary tends to the curves \rf{ro1} that touch the new limits for the critical Rossby number: ${\rm Ro}^{-}_{\rm extr}(-0.74)\approx-0.726$ and ${\rm Ro}^{+}_{\rm extr}(-0.74)\approx2.646$.}
    \label{fig4}
    \end{figure}

\subsection{Scaling law of the inductionless MRI}

What asymptotic behavior of the Reynolds and Hartmann numbers at infinity leads to maximization of negative (and simultaneously to minimization of positive) critical Rossby numbers? To get an idea, we investigate extrema of $\rm Ro$ as a solution to equation \rf{fullset}
subject to the constraints \rf{liun}. Taking, e.g., $\beta=\beta_{\rm extr}^{-}$ in equation \rf{fullset}, differentiating the result with respect to $\rm Ha$, equating it to zero and solving the equation for $\rm Re$, we find the following asymptotic relation between $\rm Ha$ and $\rm Re$ when ${\rm Ha}\rightarrow \infty$:
\be{ro6}
{\rm Re}=2{\rm Rb}\sqrt{3{\rm Rb}+2}(\sqrt{1+2{\rm Rb}}+\sqrt{2{\rm Rb}})\beta^3{\rm Ha}^3+O({\rm Ha}).
\ee
For example, at ${\rm Rb}=-1$ and $n=0$ we have $\beta=\beta_{\rm extr}^{-}=1/\sqrt{2}$. After taking this into account in \rf{ro6} we
obtain the scaling law of HMRI found in \cite{KS10}
\be{ro7}
{\rm Re}=\frac{2+\sqrt{2}}{2}{\rm Ha}^3+O({\rm Ha}).
\ee

Figure~\ref{fig4} shows the domains of the inductionless helical MRI at ${\rm Rb}=-0.74$ when the Hartmann and the Reynolds numbers are increasing in accordance with the scaling law \rf{ro6}. The instability thresholds easily penetrate the LLL and ULL as well as the Keplerian line and tend to the curves \rf{ro1} that touch the new limits for the critical Rossby number: ${\rm Ro}^{-}_{\rm extr}(-0.74)\approx-0.726$ and ${\rm Ro}^{+}_{\rm extr}(-0.74)\approx2.646$. Note that the curves \rf{ro1} correspond also to the limit of vanishing Elsasser number  $\rm N$, because according to the scaling law \rf{ro6} we have ${\rm N}\propto 1/{\rm Ha}$ as ${\rm Ha}\rightarrow \infty$. This observation makes the dispersion relation \rf{d5lb} advantageous for investigation of the inductionless versions of MRI.

\subsection{Growth rates of HMRI and AMRI and the critical Reynolds number}

We will calculate the growth rates of the inductionless MRI with the use of the dispersion relation \rf{d5lb}.
Assuming in \rf{d5lb} ${\rm Rm}:={\rm Re}{\rm Pm}=0$, we find the roots explicitly:
\be{gror6}
\lambda_{1,2} =-i n +{\rm N}\left(2\beta^2{\rm Rb}-(n\beta+1)^2\right)-\frac{1}{\rm Re}\pm2\sqrt{X+iY},
\ee
where
\be{gror6a}
X={\rm N}^2\beta^2\left(\beta^2{\rm Rb}^2+(n\beta+1)^2\right)-{\rm Ro}-1,\quad Y={\rm N}\beta({\rm Ro}+2)(n\beta+1).
\ee
Separating the real and imaginary parts of the roots, we find the growth rates of the inductionless MRI in the closed form:
\be{grorcf}
(\lambda_{1,2})_r={\rm N}\left(2\beta^2{\rm Rb}-(n\beta+1)^2\right)-\frac{1}{\rm Re}\pm\sqrt{2X+2\sqrt{X^2+Y^2}}.
\ee
Particularly, at $n=0$ we obtain the growth rates of the axisymmetric helical magnetorotational instability in the inductionless case.

Introducing the Elsasser number of the azimuthal field as
\be{na}
{\rm N}_A:=\beta^2{\rm N}
\ee and then taking the limit of $\beta \rightarrow \infty$
we obtain from the equation \rf{grorcf} the growth rates of the inductionless azimuthal MRI:
\ba{grora}
&(\lambda_{1,2})_r={\rm N}_A(2{\rm Rb}-n^2)-\frac{1}{\rm Re}\pm\sqrt{2}&\\
&\times\sqrt{{\rm N}_A^2({\rm Rb}^2+n^2)-{\rm Ro}-1+\sqrt{\left({\rm N}_A^2({\rm Rb}^2+n^2)-{\rm Ro}-1\right)^2+{\rm N}_A^2({\rm Ro}+2)^2n^2}}.&\nn
\ea
In the inviscid limit ${\rm Re}\rightarrow \infty$ the term $\frac{1}{\rm Re}$ vanishes in the expressions \rf{grorcf} and \rf{grora}.

            \begin{figure}
    \begin{center}
    \includegraphics[angle=0, width=0.85\textwidth]{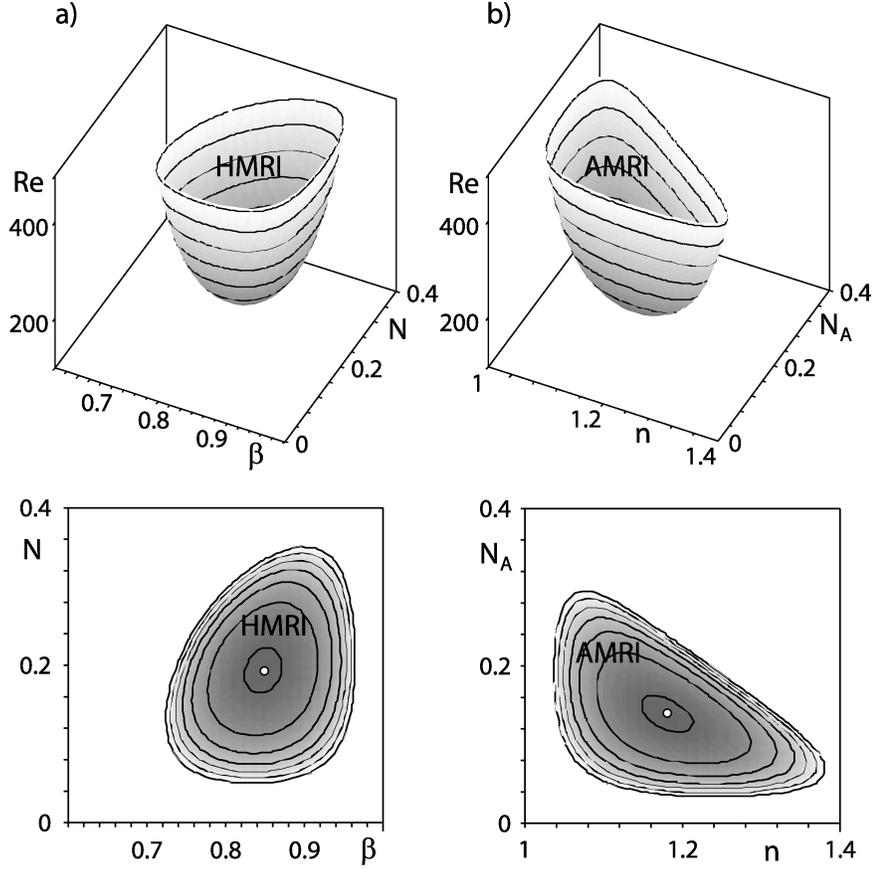}
    \end{center}
    \caption{Domain of the viscous inductionless MRI \rf{critre} of a Keplerian flow ($\rm Ro=-0.75$) at $\rm Rb=-0.75$ when (a) $n=0$ (axisymmetric HMRI) and (b) $\beta \rightarrow \infty$ (nonaxisymmetric AMRI). Open circles mark the location of the minima of the critical hydrodynamic Reynolds number: (a) $\beta \approx 0.848$, ${\rm N} \approx 0.193$, ${\rm Re} \approx 198.5$; (b) $n \approx 1.180$, ${\rm N}_A \approx 0.139$, ${\rm Re} \approx 198.5$.    }
    \label{fig5}
    \end{figure}

On the other hand, in the viscous case the condition
$(\lambda_{1,2})_r>0$ yields the critical hydrodynamic Reynolds number
beyond which inductionless MRI appears. For example, from the equation \rf{grorcf} we find the following criterion for the onset of the instability:
\be{critre}
{\rm Re}>\frac{1}{{\rm N}\left(2\beta^2{\rm Rb}-(n\beta+1)^2\right)+\sqrt{2X+2\sqrt{X^2+Y^2}}}.
\ee
With $n=0$ the criterion \rf{critre} corresponds to the axisymmetric HMRI.
A similar criterion for the onset of AMRI follows from the expression \rf{grora}.

The critical hydrodynamic Reynolds number for a Keplerian flow at ${\rm Rb}=-0.75$ is plotted in Fig.~\ref{fig5} as a function of the Elsasser number and (Fig.~\ref{fig5}(a)) $\beta$ in case of the axisymmetric HMRI and (Fig.~\ref{fig5}(b)) $n$ in case of the non-axisymmetric AMRI. In both cases the minimal $\rm Re$ at the onset of instability is about $200$.

            \begin{figure}
    \begin{center}
    \includegraphics[angle=0, width=0.95\textwidth]{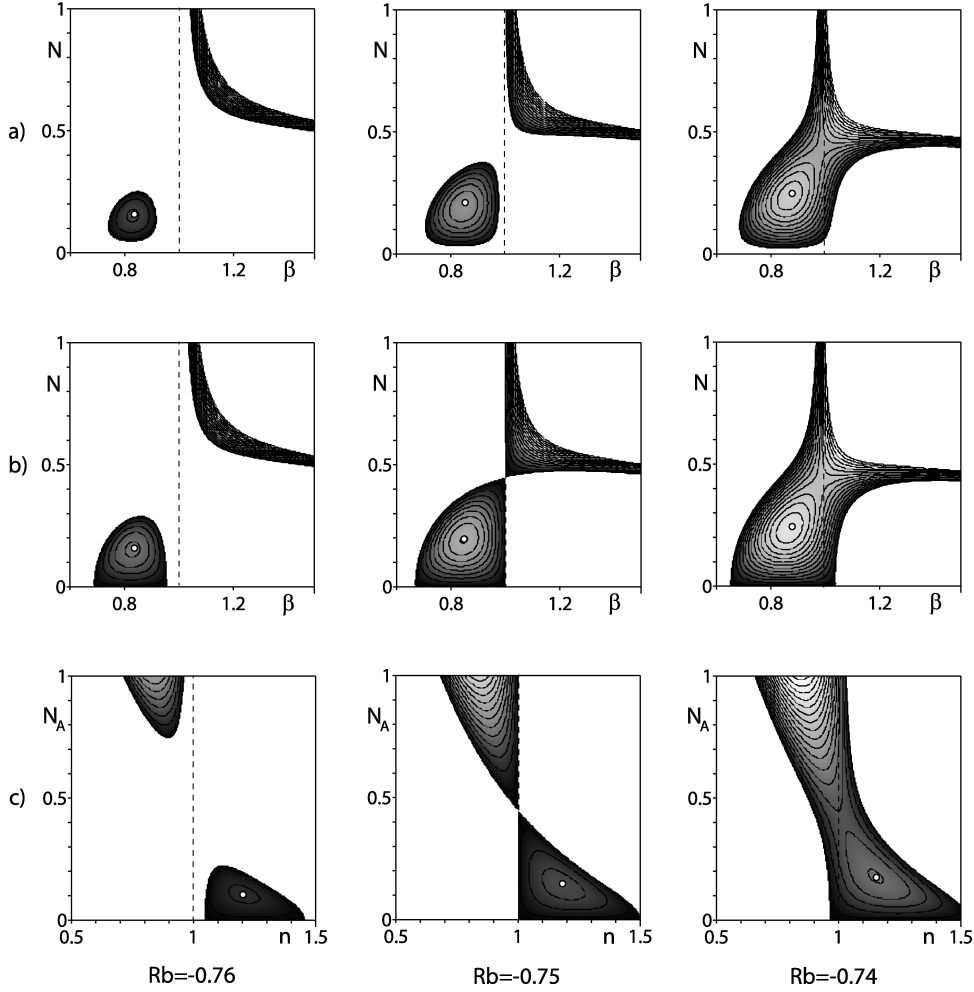}
    \end{center}
    \caption{$\rm Ro=-0.75$. Row (a): The growth rates $\lambda_r\ge 0$ of viscous (${\rm Re}=800$) HMRI ($n=0$) in the inductionless $(\rm Pm=0)$ limit in projection onto the ${\rm N}-\beta$ plane. Row (b): The same in the inviscid $({\rm Re} \rightarrow \infty)$ limit. Row (c): The growth rates $\lambda_r\ge 0$ of AMRI ($\beta\rightarrow\infty$) in the inductionless and inviscid limit in projection onto the ${\rm N}_A-n$ plane. The open circles mark the maxima of $\lambda_r$. Row (a):  $\beta \approx 0.830$, ${\rm N} \approx 0.153$, $\lambda_r \approx 0.002$; $\beta \approx 0.848$, ${\rm N} \approx 0.193$, $\lambda_r\approx0.004$;  $\beta \approx 0.869$, ${\rm N} \approx 0.236$, $\lambda_r\approx0.007$. Row (b): $\beta \approx 0.830$, ${\rm N} \approx 0.153$, $\lambda_r\approx 0.003$;  $\beta \approx 0.848$, ${\rm N} \approx 0.193$, $\lambda_r \approx   0.005$; $\beta \approx 0.869$, ${\rm N} \approx 0.236$, $\lambda_r\approx  0.008$; Row (c):    $n \approx 1.204$, ${\rm N}_A \approx 0.105 $, $\lambda_r\approx0.003$; $n \approx 1.180$, ${\rm N}_A \approx 0.139 $, $\lambda_r\approx0.005$; $n \approx 1.151$, ${\rm N}_A \approx 0.178 $, $\lambda_r\approx0.008$.}
    \label{fig6}
    \end{figure}

\subsection{HMRI and AMRI as magnetically destabilized inertial waves}

Consider the Taylor expansion of the eigenvalues \rf{gror6} with respect to the Elsasser number ${\rm N}$ in the vicinity of ${\rm N}=0$
\ba{gror7}
&\lambda_{1,2} =-i(n\mp2\sqrt{{\rm Ro}+1})-\frac{1}{\rm Re}&\nn\\
&+{\rm N}\left(-(n\beta+1)^2+2\beta^2{\rm Rb}\pm\beta\frac{({\rm Ro}+2)}{\sqrt{{\rm Ro}+1}} (n\beta+1)\right)+O({\rm N}^2).&
\ea
Expressions \rf{gror6} and \rf{gror7} generalize the result of \cite{P11} to the case of arbitrary $n$, ${\rm Re}$, and $\rm Rb$ and exactly coincide with it at $n=0$,  ${\rm Re}\rightarrow \infty$, and ${\rm Rb}=-1$.

In the absence of the magnetic field (${\rm N}=0$) the eigenvalues $\lambda_{1,2}$ correspond to damped inertial (or Kelvin) waves.
According to \rf{critre} at finite ${\rm Re}$ there exists a critical finite ${\rm N}>0$ that is necessary to trigger destabilization of the inertial waves by the magnetic field, see Fig.~\ref{fig5} and Fig.~\ref{fig6}(a). However, as the expansion \rf{gror7} demonstrates, in the limit ${\rm Re}\rightarrow \infty$ the inductionless magnetorotational instability occurs when the effect of the magnetic field is much weaker than that of the flow --- even when the Elsasser number is infinitesimally small, Fig.~\ref{fig6}(b,c).

In the inviscid case the boundary of the domain of instability \rf{critre} takes the form
\be{gror7a}
{\rm N} =\pm 2
\sqrt{
\frac{
\beta^2({\rm Ro}+2)^2(n\beta+1)^2-((n\beta+1)^2-2\beta^2{\rm Rb})^2
({\rm Ro}+1)}
{((n\beta+1)^2-4\beta^2({\rm Rb}+1))((n\beta+1)^2-2\beta^2{\rm Rb})^2(n\beta+1)^2}}.
\ee
The lines \rf{gror7a} bound the domain of non-negative growth rates of HMRI in Fig.~\ref{fig6}(b).

When ${\rm Ro}={\rm Rb}$, the stability boundary \rf{gror7a} has a self-intersection at
\be{inter1}
n=-\frac{1}{\beta}\pm2\sqrt{{\rm Rb}+1},\quad {\rm N}=\frac{\pm1}{2\beta^2}\sqrt{\frac{-(3{\rm Rb}+2)}{({\rm Rb}+1)({\rm Rb}+2)}}.
\ee
For example, when ${\rm Rb}={\rm Ro}=-0.75$ and $n=0$, the intersection happens at $\beta=1$ and ${\rm N}=\sqrt{5}/5$, Fig.~\ref{fig6}(b).
If ${\rm Rb}={\rm Ro}=-0.75$ and $\beta \rightarrow \infty$, the intersection point is at $n=1$ and ${\rm N}_A=\sqrt{5}/5$, Fig.~\ref{fig6}(c).
In general, the intersection exists at ${\rm N}\ne 0$ for
\be{gra6}
{\rm Ro}={\rm Rb}<-\frac{2}{3}.
\ee
At ${\rm Ro}={\rm Rb}
=-\frac{2}{3}$ the intersection occurs at ${\rm N}= 0$.

In Fig.~\rf{fig6} we see that when ${\rm Ro}<-\frac{2}{3}$ and ${\rm Ro}>{\rm Rb}$, the instability domain
consists of two separate regions. In the case when ${\rm Ro}<-\frac{2}{3}$ and ${\rm Ro}<{\rm Rb}$,
the two regions merge into one. When the condition \rf{gra6} is fulfilled, the two sub-domains touch each other at the point \rf{inter1}. At ${\rm Ro}={\rm Rb}=-\frac{2}{3}$ the lower region shrinks to a single point which simultaneously is the intersection point \rf{inter1} with ${\rm N}= 0$.

On the other hand, given ${\rm Ro}<-2/3$ and decreasing ${\rm Rb}$ we find that the single instability domain tends to split into two
independent regions after crossing the line ${\rm Ro}={\rm Rb}$. The further decrease in ${\rm Rb}$ yields diminishing the size of the lower  instability region, see Fig.~\ref{fig6}. At which ${\rm Rb}$ does the lower instability region completely disappear?

Clearly, the lower region disappears when the roots of the equation ${\rm N}(n) = 0$ become complex. From the expression \rf{gror7a} we derive
\be{gra8}
(n\beta +1)^2\pm\beta\frac{{\rm Ro}+2}{\sqrt{{\rm Ro}+1}}(n\beta +1)-2\beta^2{\rm Rb}=0.
\ee
The equations \rf{gra8} have the roots $n$ complex if and only if their discriminant is negative:
\be{gra9}
\frac{({\rm Ro}+2)^2}{{\rm Ro}+1}+8{\rm Rb}< 0,
\ee
which is simply the domain with the boundary given by the curve connecting the two Liu limits \rf{ro4} that is shown in Fig.~\ref{fig3}. Note that ${\rm Rb}=-2/3$ and ${\rm Ro}=-2/3$ satisfy the equation \rf{ro4},
which indicates that the line ${\rm Ro}={\rm Rb}$ is tangent to the curve \rf{ro4} at the point $(-2/3,-2/3)$ in the ${\rm Ro}$-${\rm Rb}$ plane, see Fig.~\ref{fig3}(b).

Finally, we notice that the left side of the equation \rf{gra8} is precisely the coefficient at $\rm N$ in the expansion \rf{gror7}.
In the inviscid case it determines the limit of the stability boundary as ${\rm N}\rightarrow 0$, quite in accordance with the scaling law \rf{ro6}. Solving the equation \rf{gra8} for ${\rm Ro}$, we exactly reproduce the formula \rf{ro1}. At $n=0$ and ${\rm Rb}=-1$ the equation \rf{gra8} exactly coincides with that obtained by \cite{P11}.

\section{HMRI and AMRI at very small, but finite $\rm Pm$}

An advantage of the inductionless limit discussed above is the considerable simplification of dispersion relations  at $\rm Pm=0$ or $\rm Rm=0$ that yields expressions for growth rates and stability criteria in explicit and closed form. There are, however, real physical situations that are characterized by small but finite values of the magnetic Reynolds and Prandtl numbers. Below, we demonstrate numerically that HMRI and AMRI exist also when $\rm Pm\ne0$ or $\rm Rm\ne0$. The pattern of the stability domains keeps the structure that we have found in the inductionless case. Moreover, the instability criteria of the inductionless limit serve as rather accurate guides in the physically more realistic situation of finite,  but very small, $\rm Pm$.

            \begin{figure}
    \begin{center}
    \includegraphics[angle=0, width=0.85\textwidth]{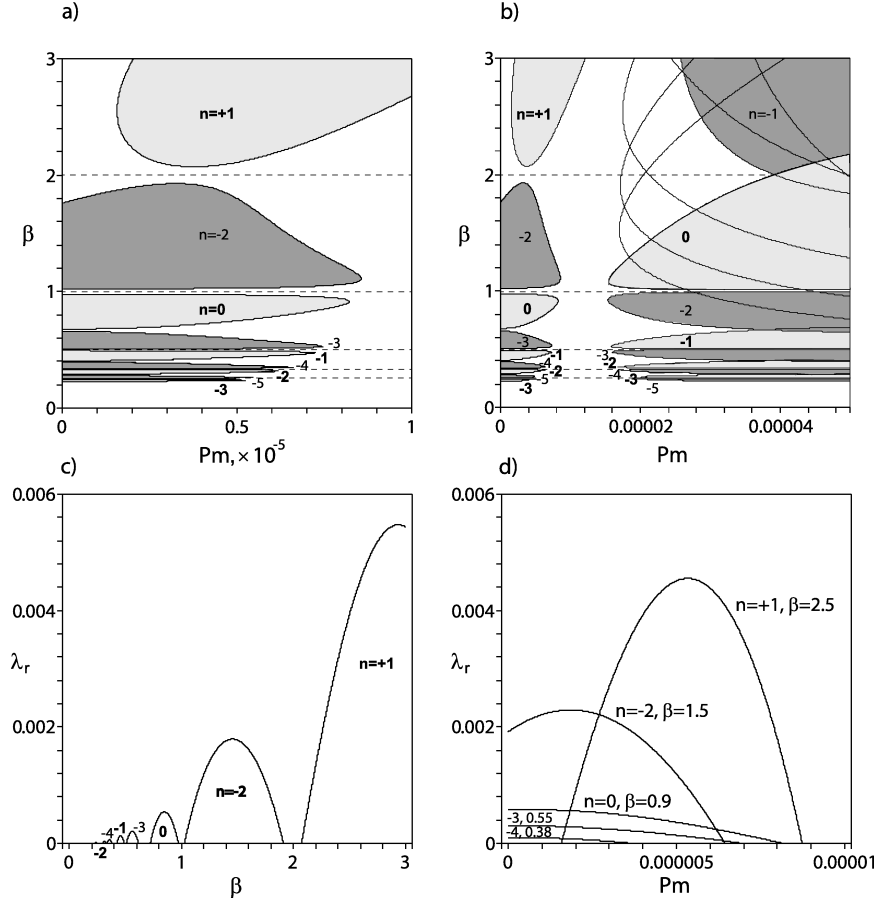}
    \end{center}
    \caption{$\rm Ha=30$, $\rm Ro=-0.75$, $\rm Re=40000$, $\rm Rb=-0.755$. (a,b) Islands of low $\rm Pm$ magnetorotational instability (axisymmetric HMRI with $n=0$, non-axisymmetric HMRI with $n\ne 0,+1$ at $\beta\le2$ and AMRI with $n=+1$ at $\beta>2$) corresponding to different modified azimuthal wave numbers $n$. (c,d) The growth rates $\lambda_r$ of the perturbation (c) as functions of $\beta$ at ${\rm Pm}=4\cdot 10^{-6}$ and different $n$ and (d) as functions of $\rm Pm$ at different $n$ and $\beta$.   }
    \label{fig4a}
    \end{figure}

\subsection{Islands of HMRI at various integer $n$ and their reconnection}

Consider the dispersion relation \rf{d4} and substitute its coefficients \rf{co1} into the Bilharz matrix \rf{bim}. Applying the Bilharz criterion \rf{bim1} to the result, we find that it is the condition of vanishing the determinant of the Bilharz matrix that determines the instability threshold. Fixing the Rossby number at the Keplerian value $\rm Ro=-0.75$ and assuming some reasonable values for the Hartmann and Reynolds numbers, e.g. $\rm Ha=30$ and $\rm Re=40000$, we choose the magnetic Rossby number slightly to the right of the line ${\rm Ro} ={\rm Rb}$, which according to the criterion \rf{ro5} should result in instability, at least in the inductionless case. For various \textit{integer} modified azimuthal wavenumbers $n$ we plot the instability domains in the $\beta-{\rm Pm}$ plane, Fig.~\ref{fig4a}(a,b).

            \begin{figure}
    \begin{center}
    \includegraphics[angle=0, width=0.99\textwidth]{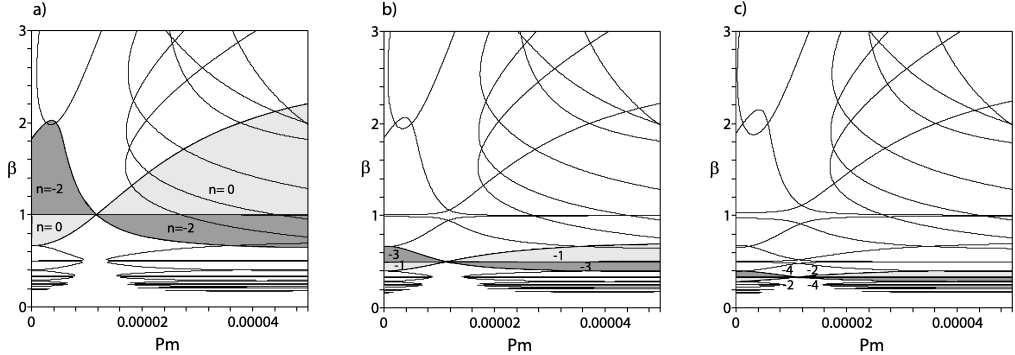}
    \end{center}
    \caption{Subsequently reconnecting instability islands at $\rm Ha=30$, $\rm Ro=-0.75$, $\rm Re=40000$ and growing ${\rm Rb}>-0.75$: (a) $\rm Rb=-0.74933$, (b) $\rm Rb=-0.747$, and (c) $\rm Rb=-0.742$.   }
    \label{fig4b}
    \end{figure}

We see that at every integer $n$ there are several instability islands (curiously resembling those originating in a pure hydrodynamical Taylor-Couette problem \citep{AHL2011}). As is visible in Fig.~\ref{fig4a}(b) they tend to group into two clusters.

The first cluster consists of the islands situated at ${\rm Pm}<10^{-5}$ and contains intervals of the positive $\beta$-axis at all $n$ but $n=+1$, Fig.~\ref{fig4a}(a).
Note that the modified azimuthal wavenumber $n$ corresponding to these islands follows exactly the sequence \rf{band3} that we have found in the inductionless case! The growth rates, i.e. the real parts of the roots $\lambda$ defined via relations \rf{d5la}, are plotted in Fig.~\ref{fig4a}(c) for a fixed value of $\rm Pm$ that cuts this cluster along a vertical line. Fig.~\ref{fig4a}(d) demonstrates the growth rates at various $\beta$ corresponding to particular values of $n$. We see that for all $n$ but $n=+1$ the growth rates tend to some positive values as ${\rm Pm}\rightarrow 0$ indicating a smooth transition to the inductionless magnetorotational instabilities that occur both for axisymmetric ($n=0$) and non-axisymmetric ($n \ne 0$) perturbations. This seems to be the main reason for the manifestation of the sequence \rf{band3} in the pattern of the instability islands of Fig.~\ref{fig4a}(c).

The second cluster of the instability islands occupies the region at ${\rm Pm}>10^{-5}$ and is encoded by the sequence \rf{band2}, see Fig.~\ref{fig4a}(b). We see that the islands of the two clusters tend to form quadruplets. Each quadruplet consists of the two pairs of islands corresponding to the indices $n$ that differ by $2$, for example: $0$ and $-2$, $-1$ and $-3$, $-2$ and $-4$ etc. Each quadruplet whose pairs are labeled with the indices $n_i,n_j\le 0$ tends to be centered at $\beta=\beta_{ij}$, where
\be{bij}
\beta_{ij}=-\frac{2}{n_i+n_j},
\ee
which is exactly the second of the equations \rf{crossing}. Moreover, the whole pattern of the instability islands in Fig.~\ref{fig4a}(b) repeats the pattern of cells in the second and third bands shown in Fig.~\ref{fig2}(a).

It is natural to ask what are the conditions for reconnection of the islands in the pairs that constitute every particular quadruplet.
To get an idea we play with the two Rossby numbers in Fig.~\ref{fig4b}. We fix ${\rm Ro}=-0.75$ and slightly increase $\rm Rb$. As a result, at ${\rm Rb}=-0.74933$ the islands with the indices $n=-2$ and $n=0$ reconnect at $\beta_{-2,0}=1$, Fig.~\ref{fig4b}(a). At ${\rm Rb}=-0.747$ these islands overlap whereas the islands in the next quadruplet with $n=-3$ and $n=-1$ reconnect at $\beta_{-3,-1}=-1/2$, Fig.~\ref{fig4b}(b). At $\rm Rb=-0.742$ the reconnection happens in the third quadruplet at $\beta_{-4,-2}=1/3$,  and so on, Fig.~\ref{fig4b}(c).

This sequence of the reconnections indicates the special role of the line ${\rm Ro}={\rm Rb}$ which seems to be even more pronounced in the inviscid limit $({\rm Re}\rightarrow \infty)$. In the following we check these hypotheses when the magnetic field has only the azimuthal component which corresponds to the limit $\beta \rightarrow \infty$.

\subsection{AMRI as a dissipation-induced instability of Chandrasekhar's equipartition solution }

In the matrix \rf{d5lc} let us replace via the relation \rf{na} the Elsasser number $\rm N$ of the axial field with the Elsasser number of the azimuthal field ${\rm N}_A$ and then let $\beta \rightarrow \infty$. If, additionally
\be{chas1}{\rm N}_A={\rm Rm}\ee
and
\be{chas2}
\rm Ro= Rb,
\ee
then in the ideal limit ($\rm Re \rightarrow \infty$, $\rm Rm \rightarrow \infty$) the roots of the dispersion relation \rf{d5lb} are
\be{chas3}
\lambda_{1,2}=0,\quad \lambda_{3,4}=-2i(n\pm1).
\ee

        \begin{figure}
    \begin{center}
    \includegraphics[angle=0, width=0.9\textwidth]{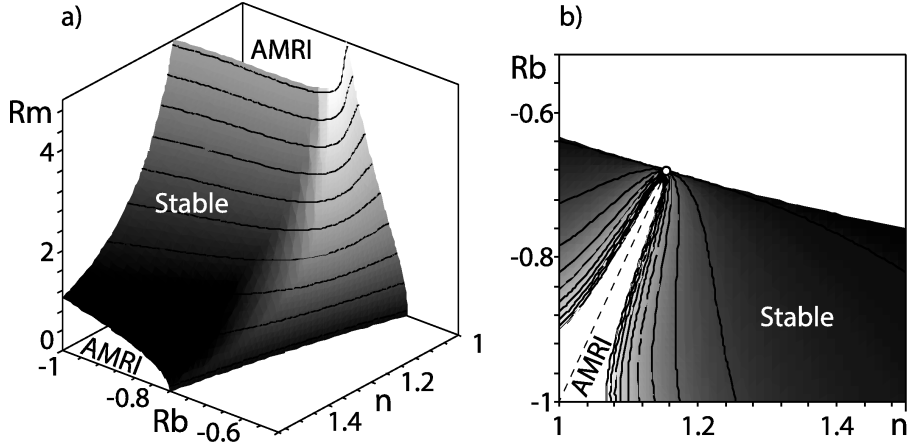}
    \end{center}
    \caption{(a) The threshold of instability \rf{cep1} at ${\rm N}_A={\rm Rm}$ and $\rm Re \rightarrow \infty$ in the $(n, {\rm Rb}, {\rm Rm})$ space and (b) its projection onto ${\rm Rb}-n$ plane.  The increase in $\rm Rm$ makes the instability domain more narrow so that  in the limit ${\rm Rm}\rightarrow \infty$ it degenerates into a ray (dashed) on the straight line \rf{cep2} that emerges from the point (open circle) with the coordinates  $n=\frac{2\sqrt{3}}{3}$ and ${\rm Rb}=-\frac{2}{3}$ and passes through the point with $n=1$ and ${\rm Rb}=-1$. }
    \label{fig1_9b}
    \end{figure}

With the use of the relations \rf{d3} it is straightforward to verify that the condition \rf{chas1} requires that $\Omega=\omega_{A_{\phi}}$. Thus, the conditions \rf{chas1} and \rf{chas2} are equivalent to \rf{ces}, which at ${\rm Rb} ={\rm Ro}=-1$ define the Chandrasekhar equipartition solution \citep{KSF2014} belonging to a wide class of exact stationary solutions of MHD equations for the case of ideal incompressible infinitely conducting fluid with total constant pressure that includes even knotted flows \citep{GK2012}. It is well-known that the Chandrasekhar equipartition solution is marginally stable \citep{Chandra56,CH61,Bog2004}. According to equation \rf{chas3} the marginal stability is preserved in the ideal case also when ${\rm Rb} ={\rm Ro}\ne -1$. Will the roots \rf{chas3} acquire only negative real parts  with the addition of electrical resistivity?

    \begin{figure}
    \begin{center}
    \includegraphics[angle=0, width=0.95\textwidth]{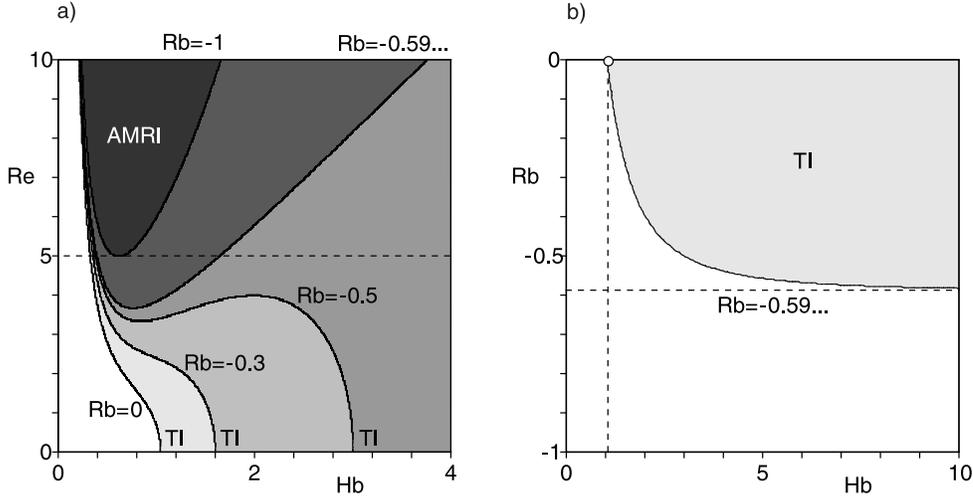}
    \end{center}
    \caption{At $n\approx1.27842$ (a) The domain of the inductionless azimuthal MRI bounded by the curve \rf{ti7} with $\rm Ro=-1$ and $\rm Rb=-1$ that gradually transforms into the domain that includes the  Tayler instability (TI) as $\rm Rb$ increases from $-1$ to $0$; (b) Tayler instability corresponding to $\rm Ro=0$ and $\rm Re=0$. The open circle marks the onset of the standard TI at $\rm Rb=0$ with the critical ${\rm Hb}\approx1.04117$.}
    \label{fig1_9c}
    \end{figure}

In general, the answer is no. Indeed, under the constraints \rf{chas1} and \rf{chas2} in the limit of vanishing viscosity ($\rm Re \rightarrow \infty$) the Bilharz criterion applied to the dispersion relation \rf{d5lb} gives the following threshold of instability:
\ba{cep1}
    &16(n^2-{\rm Rb}^2)(n^2-{\rm Rb}-2)^2{\rm Rm}^4&\nn\\
    &+(n^6-12n^2{\rm Rb}^2+32n^2({\rm Rb}+1)-16{\rm Rb}^2({\rm Rb}+2)){\rm Rm}^2&\nn\\
    &-4{\rm Rb}^2+4n^2({\rm Rb}+1)=0.&
   \ea

In the $(n, {\rm Rb}, {\rm Rm})$ space the domain of instability is below the surface specified by equation \rf{cep1}, see Fig.~\ref{fig1_9b}(a). When the electrical resistivity is vanishing, the cross-sections of the instability domain in the ${\rm Rb}-n$ plane become smaller and in the limit ${\rm Rm}\rightarrow \infty$ they tend to the ray on the line
\be{cep2}
    {\rm Rb}=-\frac{2}{3}+\frac{n\sqrt{3}-2}{3(2-\sqrt{3})}
\ee
that starts at the point with the coordinates  $n=\frac{2\sqrt{3}}{3}$ and ${\rm Rb}=-\frac{2}{3}$ and passes through the point with $n=1$ and ${\rm Rb}=-1$, Fig.~\ref{fig1_9b}(b). For example, at $\rm Ro=Rb=-\frac{3}{4}$ the equation \rf{cep2} yields
\be{cep2a}
n=\frac{1}{4}+\frac{\sqrt{3}}{2}.
\ee

 On the contrary, when the magnetic Reynolds number ${\rm Rm}$ decreases, the instability domain widens up and in the inductionless limit at $\rm Rm =0$ it is bounded by the curve
\be{cep3}
{\rm Rb}=\frac{n(n-\sqrt{n^2+4})}{2}.
\ee

The wide part of the instability domain shown in Fig.~\ref{fig1_9b}(a) that exists at small $\rm Rm$ belongs to the realm of the azimuthal magnetorotational instability (AMRI). We see that this instability quickly disappears with the increase of $\rm Rm$  or $\rm Rb$. On the other hand, the ideal solution with the roots \rf{chas3} that corresponds to the limit ${\rm Rm}\rightarrow \infty$ is destabilized by the electrical resistivity. For $n$ given by equation \rf{cep2a} we have, for example, an unstable root $\lambda\approx 0.00026 - i0.00493$ at $\rm Rm=100$. In general, if $\rm Rb$ and $n$ satisfy \rf{cep2}, then already an \textit{infinitesimally weak} electrical resistivity destabilizes the solution specified by the constraints \rf{chas1} and \rf{chas2} at vanishing kinematic viscosity that includes Chandrasekhar's equipartition solution as a special case.
This \textit{dissipation-induced instability} \citep{Ki06,KV10,K2013dg} further develops into the AMRI with ${\rm Rm}$ decreasing to zero.

\section{Transition from AMRI to the Tayler instability}

The Tayler instability \citep{TAYLER,RS10} is a current-driven,
kink-type instability that taps into the magnetic field
energy of the electrical current in the fluid.
Although its plasma-physics counterpart has been known
for a long time, its occurrence in a liquid metal
was observed only recently \citep{SEILMAYER2012}.
In the context of the on-going liquid-metal experiments in
the frames of the DRESDYN project \citep{DRESDYN}
it is interesting to get an insight on the transition between
the azimuthal magnetorotational instability and the Tayler instability.

    \begin{figure}
    \begin{center}
    \includegraphics[angle=0, width=0.9\textwidth]{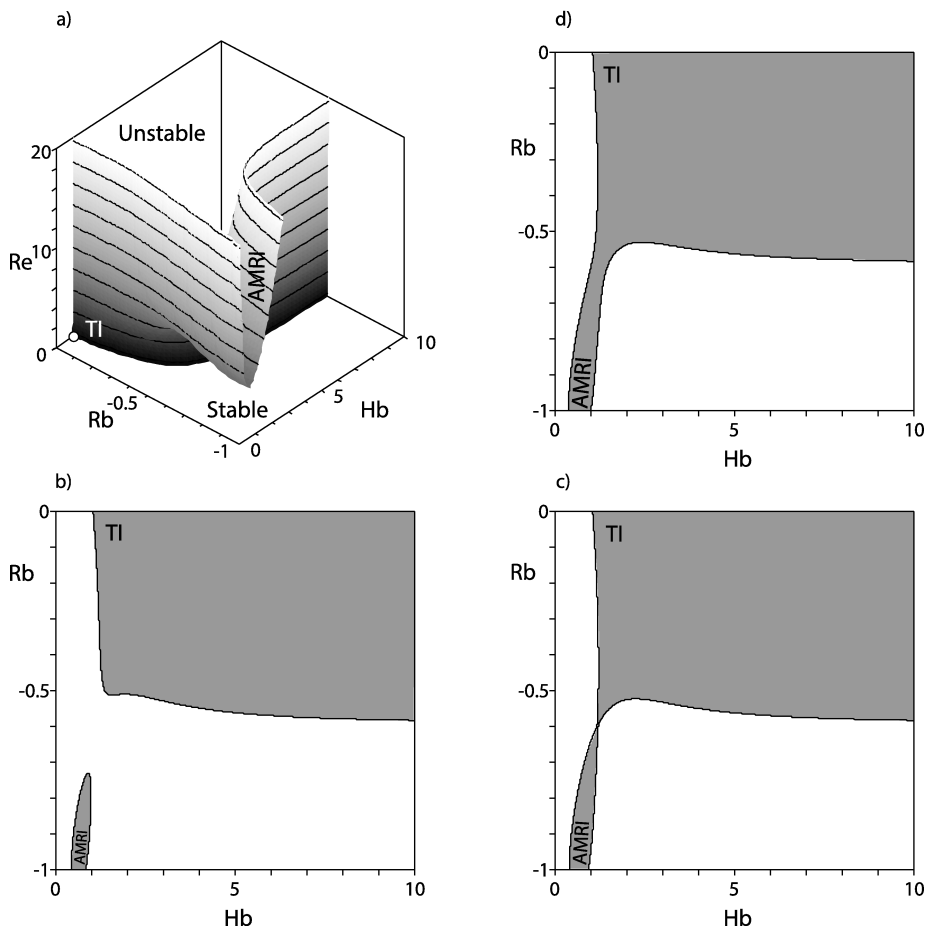}
    \end{center}
    \caption{Under the assumption that ${\rm Ro}({\rm Rb})=-\sqrt{-\rm Rb^2-2\rm Rb}$
    and $n\approx1.27842$ (a) The domain of the \textit{inductionless} instability
    bounded by the surface \rf{ti7} in the $({\rm Hb},{\rm Rb},{\rm Re})$ space and
    its cross-sections at (b) $\rm Re=5.4$, (c) $\rm Re=5.734$, and (d) $\rm Re=6$.
    The domains of TI and AMRI reconnect via a saddle point at $\rm Re=5.734$.}
    \label{fig1_9d}
    \end{figure}

Consider the instability threshold \rf{fullset} obtained in the inductionless approximation (${\rm Pm}=0$).
Let us introduce the Hartmann number corresponding to the pure azimuthal magnetic field as
\be{hb1}
{\rm Hb}:=\beta {\rm Ha},
\ee
so that ${\rm N}_A=\frac{{\rm Hb}^2}{\rm Re}$.
Substituting \rf{hb1} into  \rf{fullset} and then letting $\beta\rightarrow \infty$, we find
\be{ti7}
{\rm Re}^2=\frac{((1+{\rm Hb}^2n^2)^2-4{\rm Hb}^2{\rm Rb}(1+{\rm Hb}^2n^2)-4{\rm Hb}^4n^2)(1+{\rm Hb}^2(n^2-2{\rm Rb}))^2}{4({\rm Hb}^4{\rm Ro}^2n^2-((1+{\rm Hb}^2(n^2-2{\rm Rb}))^2-4{\rm Hb}^4n^2)({\rm Ro}+1))}.
\ee
Note that the expression \rf{ti7} can also be derived  from the equation \rf{grora}.

    \begin{figure}
    \begin{center}
    \includegraphics[angle=0, width=0.9\textwidth]{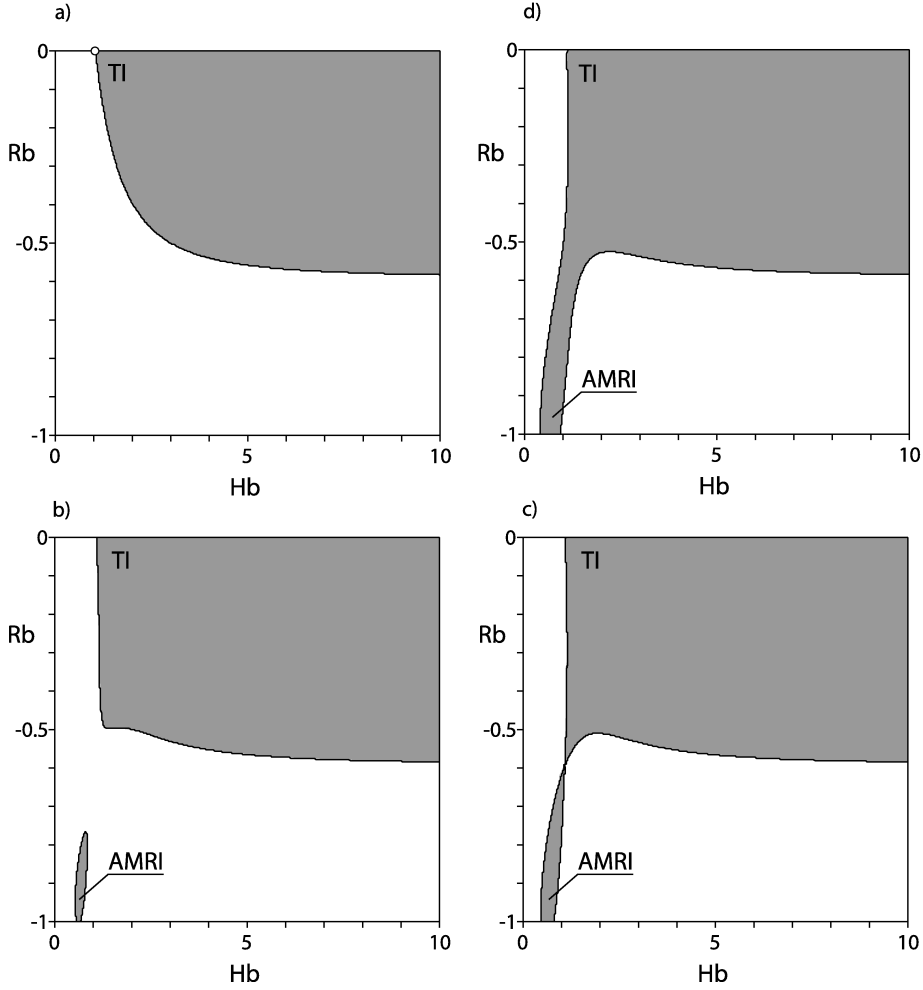}
    \end{center}
    \caption{The instability domain in the assumption that magnetic Prandtl number takes a finite value $\rm Pm=0.05$, $n = 1.27842$, and $\rm Ro=-\sqrt{\rm -Rb^2-2Rb}$  at (a) $\rm Re=0$, (b) $\rm Re=3.8$, (c) $\rm Re=3.958$, (d) $\rm Re=4.2$. }
    \label{fig1_9e}
    \end{figure}

Consider the threshold of instability \rf{ti7} in the two special cases corresponding to the lower left and the upper right corners of the ${\rm Ro}-{\rm Rb}$ diagram shown in Fig.~\ref{fig3}(b). At $\rm Ro=-1$ and $\rm Rb=-1$ the function ${\rm Re}(n,{\rm Hb})$ that bounds the domain of AMRI has a minimum ${\rm Re}\approx4.99083$ at $n\approx1.27842$ and ${\rm Hb}\approx0.61185$, see Fig.~\ref{fig1_9c}(a).

Putting $\rm Re=0$ in \rf{ti7}, we find the threshold for the critical azimuthal magnetic field
\be{ti3}
{\rm Rb}= \frac{({\rm Hb}^2n^2+1)^2-4{\rm Hb}^4n^2}{4{\rm Hb}^2(1+{\rm Hb}^2n^2)}
\ee
that destabilizes electrically conducting fluid at rest (cf. criterion \rf{tay1om}). At ${\rm Rb}=0$ expression \rf{ti3} gives the value of the azimuthal magnetic field
at the onset of the standard Tayler instability (cf. criterion \rf{tay2om})
\be{ti5}
{\rm Hb}=\frac{1}{\sqrt{1-(1- |n|)^2}}.
\ee
For example, at $n\approx1.27842$
\be{ti6}
{\rm Hb}\approx1.04117,
\ee
see Fig.~\ref{fig1_9c}(b). In the following, we prefer to extend the notion of the Tayler instability (TI) to the whole domain bounded by the curve \rf{ti3} and shown in gray in Fig.~\ref{fig1_9c}(b).

    \begin{figure}
    \begin{center}
    \includegraphics[angle=0, width=0.9\textwidth]{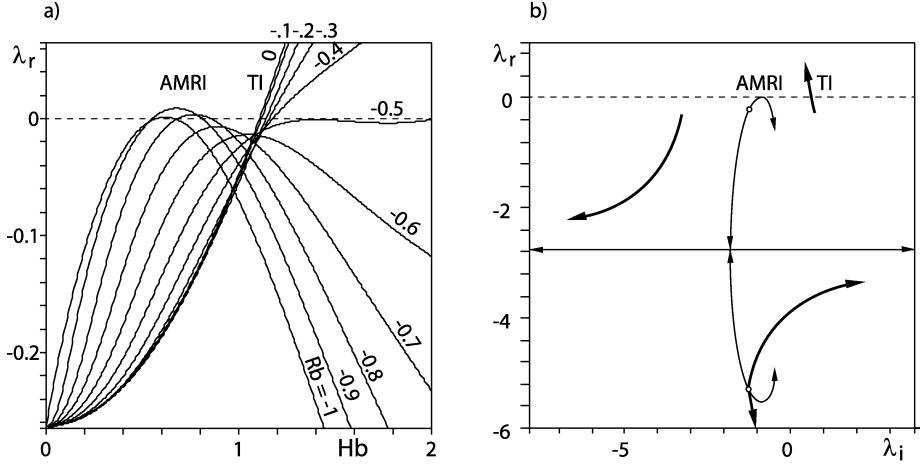}
    \end{center}
    \caption{At $\rm Pm=0.05$, $n = 1.27842$, $\rm Re=3.8$, and $\rm Ro=-\sqrt{\rm -Rb^2-2Rb}$ (a) the growth rates $\lambda_r({\rm Hb})$ as $\rm Rb$ changes from $-1$ to $0$; (b) the movement of the roots in the complex plane for (fine lines) ${\rm Rb}=-1$ and (bold lines) ${\rm Rb}=0$. }
    \label{fig1_9f}
    \end{figure}

How the domains of the Tayler instability and AMRI are related to each other? Is there a connection between them in the parameter space?

As is seen in Fig.~\ref{fig1_9c}(b), with the decrease in $\rm Rb$, the critical value of $\rm Hb$ at the onset of TI
\be{onsetti}
{\rm Hb}=\frac{1}{\sqrt{2{\rm Rb}+2\sqrt{{\rm Rb}^2+n^2}-n^2}}.
\ee
increases and, as follows from Eqs.~\rf{ti3} and \rf{onsetti}, tends to infinity as ${\rm Rb}\rightarrow \frac{n^2}{4}-1$. The asymptote \be{ati}{\rm Rb}=\frac{n^2}{4}-1\ee is shown in Fig.~\ref{fig1_9c}(b) as a horizontal dashed line. If we look at the instability domain in the $\rm Re$-$\rm Hb$ plane at the fixed $\rm Ro$ and $n$, then we will see how the domain of the pure AMRI gradually transforms into that containing the Tayler instability when $\rm Rb$ varies from $-1$ to $0$. The boundary of TI at $\rm Re=0$ originates exactly at the moment when ${\rm Rb}=\frac{n^2}{4}-1$, Fig.~\ref{fig1_9c}(a).

Let us look at  the ${\rm Ro}-{\rm Rb}$ diagram shown in Fig.~\ref{fig3}(b). To connect the two opposite corners of it we obviously need to take a path that lies below the line $\rm Ro = Rb$. Indeed, any path above this line penetrates the limiting curve \rf{ro4} which creates an obstacle for connecting the two regions shown in Fig.~\ref{fig1_9c}. On the contrary, any path below the diagonal in Fig.~\ref{fig3}(b) lies within the instability domain which opens a possibility to connect the regions of AMRI and standard TI corresponding to $\rm Rb=0$ and $\rm Ro=0$. Therefore, it is the dependency ${\rm Ro}({\rm Rb})$ that controls the transition from the azimuthal magnetorotational instability to the Tayler instability.

Assume, for example, that $\rm Ro=-\sqrt{-\rm Rb^2-2\rm Rb}$, which is a part of the unit circle that  connects the point with $\rm Rb=-1$ and $\rm Ro=-1$ and the point with $\rm Rb=0$ and $\rm Ro=0$ in Fig.~\ref{fig3}(b). Substituting this dependency into the expression \rf{ti7}, we can plot the domain of instability in the $({\rm Hb},{\rm Rb},{\rm Re})$ space for a given $n$, Fig.~\ref{fig1_9d}(a). The cross-section of the domain at $\rm Re=0$ yields exactly the region of the Tayler instability shown in Fig.~\ref{fig1_9c}(b). The cross-section at $\rm Rb=-1$ is the domain of AMRI in Fig.~\ref{fig1_9c}(a).

At $\rm Re < 4.99083...$ the azimuthal MRI with $n\approx1.27842$ at ${\rm Rb}=-1$ does not exist. With the increase in the Reynolds number the island of AMRI originates, Fig.~\ref{fig1_9d}(b), that further reconnects with the viscosity-modified domain of TI at a saddle point, Fig.~\ref{fig1_9d}(c). At higher values of $\rm Re$ the two domains merge into one, Fig.~\ref{fig1_9d}(d).

What happens with the inductionless instability diagram of
Fig.~\ref{fig1_9d} when the magnetic Prandtl number takes finite values?
It turns out that at $\rm Pm \ll 1$ the difference is very
small and is not qualitative. Moreover, at $\rm Re=0$  the
domain of the Tayler instability is still given by
equation \rf{ti3} and thus coincides with that shown
in Fig.~\ref{fig1_9c}(b).  Nevertheless, the critical
Reynolds number at the saddle point and at the onset of
AMRI slightly decreases with the increase in $\rm Pm$, Fig.~\ref{fig1_9e}.

Finally, we plot in Fig.~\ref{fig1_9f}(a) the growth
rates $\lambda_r(\rm Hb)$ as $\rm Rb$ varies from $-1$ to
$0$. One can see that the growth rates of the Tayler
instability monotonously increase and become positive
at $\rm Hb > 1$. When $\rm Rb$ is smaller than about $-0.5$,
the function $\lambda_r(\rm Hb)$ has a maximum that can
both  lie below zero and exceed it in dependence on $\rm Rb$.
The latter weak instability is the azimuthal magnetorotational instability.

In Fig.~\ref{fig1_9f}(b) the movement of the roots corresponding
to TI and AMRI is presented. There are indications that the origin
of the both instabilities might be related to the splitting of a
multiple zero root. Note that in the inductionless case the
detailed analysis of the roots movement and their bifurcation
is possible with the use of equations \rf{gror6} and \rf{grora}.
We leave this for a future work.

\section{Conclusion}

Using the formal short-wavelength asymptotic expansions of geometric optics, we have derived the local transport PDEs for the amplitude of the localized non-axisymmetric perturbation  of a rotating flow under the influence of a constant vertical and
an azimuthal magnetic field with
arbitrary radial dependence. Looking for the solution of the local transport PDEs in the modal form we have derived a  dispersion relation of AMRI that is suitable for testing stability in the case of both ideal and dissipative MHD. The dispersion relation of HMRI in viscous and resistive setup is derived separately by applying the WKB approximation in the radial direction only.
Particularly, in the limit of vanishing
magnetic Prandtl number as well as for small, but finite $\rm Pm$, we have shown that Keplerian profiles
can well be destabilized by HMRI or AMRI if only the azimuthal
field profile is slightly flatter than $1/R$. We have also
shown that the line where
the hydrodynamic and the magnetic Rossby number are equal, plays an
essential role for the connectedness of the instability domain. We have found that the marginally stable solution corresponding to $\rm Ro=Rb$ and ${\rm N}_A={\rm Rm}$ in the limit when both ${\rm Re} \rightarrow \infty$ and ${\rm Rm} \rightarrow \infty$ is generically destabilized by weak but finite electrical resistivity  (and in particular cases already by infinitesimally weak electrical resistivity) resulting in AMRI, which is therefore a dissipation-induced instability. In particular, when ${\rm N}_A={\rm Rm}$ and $\rm Ro=Rb=-1$, AMRI is a dissipation-induced instability of the Chandrasekhar equipartition solution of ideal MHD. Finally, we have found an analytical condition on $ \rm Rb$ separating the regimes of pure AMRI and Tayler instability.

With view on astrophysical applications one has definitely to note that
any flatter than $1/R$ profile of $B_{\phi}$ would require some
finite magnetic Reynolds number for the necessary
induction effects to occur.
Nevertheless, our results still provide a real extension of the
applicability of MRI, since the Lundquist numbers are allowed to
be arbitrarily small, although the growth rate, which is
proportional to the interaction parameter, would then be
rather small.

The consequences of our findings for those parts of accretion disks
with small magnetic Prandtl numbers are still to be elaborated.
The action of MRI in the dead zones of protoplanetary disks
is an example for which the extended parameter region might have
consequences.
Particular attention  should also be given to the
possibility of quasi-periodic oscillations which might easily result from
the sensitive dependence of the action of HMRI on the radial profile of
of $B_{\phi}$ and the ratio of the latter to $B_z$. We notice however that pure hydrodynamical scenarios of transition to turbulence in the dead zones had also been proposed \citep{Marcus}.

As for liquid metal experiments, our results give strong impetus
for a special set-up
in which the magnetic Rossby number can be adjusted by using two
independent electrical currents, one through an central, insulated rod,
the second one through the liquid metal. A liquid sodium experiment
dedicated exactly to this problem
is presently being designed in the framework of the DRESDYN
project \citep{DRESDYN}. Apart from this, the recently
observed, and numerically confirmed, strong sensitivity
of AMRI on a slight symmetry breaking of an external
magnetic field \citep{SEILMAYER2013} may also be related to
our findings.

\section*{Acknowledgement}
This work was supported by Helmholtz-Gemeinschaft Deutscher
Forschungszentren (HGF) in frame of the Helmholtz Alliance LIMTECH,
as well as by Deutsche
Forschungsgemeinschaft in frame of the SPP 1488 (PlanetMag). F.S.
acknowledges fruitful discussions with Marcus Gellert,
Rainer Hollerbach, and G\"unther R\"udiger. We thank an anonymous referee for helpful comments.

\appendix

\section{Connection to the work by \cite{FV95}}\label{appD}

\cite{FV95} considered stability with respect to non-axisymmetric perturbation of the general toroidal ideal MHD equilibrium
with $\boldsymbol{u}_0=R\Omega(R,z)\boldsymbol{e}_{\phi}$ and $\boldsymbol{B}_0=B^0_{\phi}(R,z)\boldsymbol{e}_{\phi}=R\sqrt{\rho \mu_0}A(R,z)\boldsymbol{e}_{\phi}$, where the equilibrium equation requires
$\partial_z(\Omega^2-A^2)=0$. If we neglect the $z$-dependence
and assume that the artificially introduced  quantization constant $\tau=0$, their dispersion relation derived in the short-wavelength approximation is given by the characteristic polynomial of the matrix
$$
{\bf F}=\left(
          \begin{array}{cccc}
            -i m (\Omega-A)  & \alpha^2(\Omega -A)& 0 & \alpha^2(\Omega +A) \\
            A-\Omega & -im(\Omega-A) & -A-\Omega-R\partial_R(\Omega+A)  & 0 \\
            0 & \alpha^2(\Omega -A) & -i m (\Omega+A)  & \alpha^2(\Omega +A) \\
            A-\Omega-R\partial_R(\Omega-A) & 0 & -(\Omega+A)  & -im(\Omega+A) \\
          \end{array}
        \right),
$$
where $\alpha=\frac{k_z}{\sqrt{k_z^2+k_R^2}}$, $A=\omega_{A_{\phi}}$, and according to Eq.~\rf{hro} and Eq.~\rf{freqa}, $R\partial_R\Omega=2\Omega{\rm Ro}$ and $R\partial_R A=2\omega_{A_{\phi}}{\rm Rb}$.
In the new notation the coefficients of the dispersion relation derived by
\cite{FV95} exactly coincide with that given by Eq.~\rf{cp2} were one has to set $\omega_{\nu}=0$, $\omega_{\eta}=0$, and $\omega_{A_z}=0$.

To show this, we add/subtract the second of the amplitude equations \rf{g13} divided by $\sqrt{\rho\mu_0}$ to/from the
first one in order to write it via the Elsasser variables $\boldsymbol{a}_{\pm}:=\boldsymbol{u}^{(0)}\pm\frac{\boldsymbol{B}^{(0)}}{\sqrt{\rho \mu_0}}$
\ba{fv1a}
\dot{\boldsymbol{a}}_+&=&-\left(\left[ \boldsymbol{u}_0-\frac{\boldsymbol{B}_0}{\sqrt{\rho\mu_0}} \right]\cdot \nabla\right)\boldsymbol{a}_+-\left(\mathcal{U}{+}\frac{\mathcal{B}}{\sqrt{\rho \mu_0}} \right)\boldsymbol{a}_--\frac{\omega_{\nu} +\omega_{\eta}}{2} \boldsymbol{a}_+-\frac{\omega_{\nu} -\omega_{\eta}}{2} \boldsymbol{a}_-\nn\\
&+&\frac{\boldsymbol{k}\boldsymbol{k}^T}{|\boldsymbol{k}|^2}\left(\left[\mathcal{U}{+}\frac{\mathcal{B}}{\sqrt{\rho \mu_0}}\right]\boldsymbol{a}_-{+}\left[\mathcal{U}{-}\frac{\mathcal{B}}{\sqrt{\rho \mu_0}}\right]\boldsymbol{a}_+ \right){+}\frac{\boldsymbol{k}\boldsymbol{k}^T}{|\boldsymbol{k}|^2}\left(\frac{\mathcal{B}}{\sqrt{\rho \mu_0}}{-}\left(\frac{\boldsymbol{B}_0}{\sqrt{\rho \mu_0}}\cdot \nabla\right) \right)\frac{\boldsymbol{a}_+{-}\boldsymbol{a}_-}{2},\nn
\ea
\ba{fv1b}
\dot{\boldsymbol{a}}_-&=&-\left(\left[ \boldsymbol{u}_0+\frac{\boldsymbol{B}_0}{\sqrt{\rho\mu_0}} \right]\cdot \nabla\right)\boldsymbol{a}_--\left(\mathcal{U}{-}\frac{\mathcal{B}}{\sqrt{\rho \mu_0}} \right)\boldsymbol{a}_+-\frac{\omega_{\nu} +\omega_{\eta}}{2} \boldsymbol{a}_--\frac{\omega_{\nu} -\omega_{\eta}}{2} \boldsymbol{a}_+\nn\\
&+&\frac{\boldsymbol{k}\boldsymbol{k}^T}{|\boldsymbol{k}|^2}\left(\left[\mathcal{U}{+}\frac{\mathcal{B}}{\sqrt{\rho \mu_0}}\right]\boldsymbol{a}_-{+}\left[\mathcal{U}{-}\frac{\mathcal{B}}{\sqrt{\rho \mu_0}}\right]\boldsymbol{a}_+ \right){+}\frac{\boldsymbol{k}\boldsymbol{k}^T}{|\boldsymbol{k}|^2}\left(\frac{\mathcal{B}}{\sqrt{\rho \mu_0}}{-}\left(\frac{\boldsymbol{B}_0}{\sqrt{\rho \mu_0}}\cdot \nabla\right) \right)\frac{\boldsymbol{a}_+{-}\boldsymbol{a}_-}{2}.\nn
\ea
Looking for the solution $\boldsymbol{a}_{\pm}=\widehat{\boldsymbol{a}}_{\pm}e^{\gamma t +im\phi}$, taking into account that $k_R=k_z\sqrt{1-\alpha^2}\alpha^{-1}$, and eliminating the $z$-components of the vectors $\boldsymbol{a}_{\pm}$ via the constraints $\boldsymbol{a}_{\pm}\cdot \boldsymbol{k}=0$ following from Eq.~\rf{g4} we arrive at the equation ${\bf F}\widehat{\bf b}=\gamma \widehat{\bf b}$, where $\widehat{\bf b}=(\widehat{ a}_{+,R},\widehat{ a}_{+,\phi},\widehat{ a}_{-,R},\widehat{a}_{-,\phi})^T$ and $\bf F$ is the matrix defined above in the case when $\omega_{\nu}=0$ and $\omega_{\eta}=0$.

\section{Coefficients of the dispersion relations}\label{appC}

The complex characteristic polynomial \rf{dr2} of the matrix \rf{mri11} has the following coefficients:
\ba{cp2}
&a_0=1,\quad b_0=0,\quad
a_1 = 2(\omega_{\eta}+\omega_{\nu}), \quad b_1 = 4m\Omega,&\nn\\
&a_2 = -4\alpha^2\omega_{A_{\phi}}^2{\rm Rb}+4\alpha^2\Omega^2(1+{\rm Ro})+(\omega_{\nu}+\omega_{\eta})^2-2(3m^2\Omega^2-(m\omega_{A_{\phi}}+\omega_{A_z})^2-\omega_{\nu}\omega_{\eta}),&\nn\\
&b_2 = \frac{3}{4}a_1b_1,&\nn\\
&a_3 = 2(\omega_{\nu}+\omega_{\eta})(-2\alpha^2\omega_{A_{\phi}}^2{\rm Rb}-(3m^2\Omega^2-(m\omega_{A_{\phi}}+\omega_{A_z})^2-\omega_{\nu}\omega_{\eta}))+8\alpha^2\Omega^2\omega_{\eta}(1+{\rm Ro}),&\nn\\
&b_3 = \frac{a_2b_1}{2}+\frac{b_1^3}{8}-8\Omega\alpha^2(\omega_{A_z}+m\omega_{A_{\phi}})\omega_{A_{\phi}},&\nn\\
&a_4=((4\Omega^2(\omega_{A_z}+m\omega_{A_{\phi}})^2-4\Omega^4m^2+4\Omega^2\omega_{\eta}^2){\rm Ro}+8\omega_{A_{\phi}}\omega_{A_z} m\Omega^2+4\Omega^2\omega_{\eta}^2-4\Omega^4m^2)\alpha^2&
\nn\\
&-\Omega^2m^2(\omega_{\nu}+\omega_{\eta})^2+(m^2\Omega^2-(\omega_{A_z}+m\omega_{A_{\phi}})^2-\omega_{\nu}\omega_{\eta})^2&\nn\\
&+
4({\rm Rb}+1)\omega_{A_{\phi}}^2\alpha^2(m^2\Omega^2-(\omega_{A_z}+m\omega_{A_{\phi}})^2)+4\omega_{A_{\phi}}^2\alpha^2(m^2\Omega^2-\omega_{\nu}\omega_{\eta}{\rm Rb}),&\nn\\
&b_4 = -4m\Omega\alpha^2[({\rm Ro}+1)(\omega_{\eta}-\omega_{\nu})+({\rm Rb}+1)(\omega_{\eta}+\omega_{\nu})]\omega_{A_{\phi}}^2&\nn\\
&-4\omega_{A_z}\Omega\alpha^2({\rm Ro}(\omega_{\eta}-\omega_{\nu})+ 2\omega_{\eta})\omega_{A_{\phi}}&\nn\\
&+2m\Omega(4\alpha^2\Omega^2\omega_{\eta}({\rm Ro}+1)+(\omega_{\nu}+\omega_{\eta})(\omega_{\nu}\omega_{\eta}-m^2\Omega^2+(\omega_{A_z}+m\omega_{A_{\phi}})^2)).&
\ea

The coefficients of the complex polynomial \rf{d4} are:
\ba{co1}
&a_0=1,\quad b_0=0,\quad
a_1=2\left(\sqrt{\rm Pm}+\frac{1}{\sqrt{\rm Pm}}\right),\quad b_1=4 n {\rm Re} \sqrt{\rm Pm},&\nn \\
&a_2=2(\beta^2{\rm Ha}^2-3{\rm Re}^2{\rm Pm})n^2+4\beta{\rm Ha}^2 n&\nn\\
   &+2(1+(1-2{\rm Rb}\beta^2){\rm Ha}^2)+4{\rm Re}^2(1+{\rm Ro}){\rm Pm}+\frac{a_1^2}{4},\quad
b_2= 6n{\rm Re}(1+{\rm Pm}),&\nn\\
&a_3= a_1(\beta^2{\rm Ha}^2-3{\rm Re}^2{\rm Pm})n^2+2a_1\beta {\rm Ha}^2 n&\nn\\
   &+a_1(1+(1-2{\rm Rb}\beta^2){\rm Ha}^2)+8{\rm Re}^2(1+{\rm Ro})\sqrt{\rm Pm},&\nn\\
&b_3= 4n^3\sqrt{\rm Pm}{\rm Re}(\beta^2{\rm Ha}^2-{\rm Re}^2{\rm Pm})&\nn\\
   &+2n{\rm Re}(4{\rm Pm}^2{\rm Re}^2(1+{\rm Ro})+(1+{\rm Pm})^2+2{\rm Pm}(1+{\rm Ha}^2))/\sqrt{\rm Pm}&\nn\\
   &-8(1-n^2+\beta n(1+{\rm Rb}))\beta{\rm Ha}^2{\rm Re}\sqrt{\rm Pm},&\nn\\
&a_4=((\beta^2{\rm Ha}^2-{\rm Re}^2{\rm Pm})n^2+2{\rm Ha}^2\beta n+{\rm Ha}^2+2{\rm Pm}{\rm Re}^2)^2&\nn\\
&+2(2{\rm Re}^2{\rm Pm}{\rm Ro}+1)(({\rm Ha}^2\beta^2-{\rm Re}^2{\rm Pm})n^2+2{\rm Ha}^2\beta n+{\rm Ha}^2)-(1+{\rm Pm})^2{\rm Re}^2n^2&\nn\\
&+4{\rm Re}^2(1+{\rm Ro})-({\rm Ha}^2+2{\rm Pm}{\rm Re}^2)^2+{\rm Ha}^4+1-4{\rm Rb}\beta^2{\rm Ha}^2&\nn\\
&-4{\rm Ha}^2\beta^2({\rm Ha}^2(\beta n+1)^2-{\rm Pm}{\rm Re}^2 n^2)(1+{\rm Rb}),&\nn\\
&b_4= 2{\rm Re}(1+{\rm Pm})(\beta^2{\rm Ha}^2-{\rm Re}^2{\rm Pm})n^3+4{\rm Re}{\rm Ha}^2\beta(1+{\rm Pm})n^2&\nn\\
&+2{\rm Re}(2(1+{\rm Ro})(2{\rm Re}^2{\rm Pm}-\beta^2{\rm Ha}^2(1{-}{\rm Pm})){+}(1{+}{\rm Pm})(1+{\rm Ha}^2(1-2\beta^2(1+{\rm Rb}))))n&\nn\\
&-4\beta{\rm Ha}^2{\rm Re}(2+(1-{\rm Pm}){\rm Ro}).&
\ea

The next equation gives the coefficients of the dispersion relation \rf{d5lb}:
\ba{co2}
&a_0=1,\quad b_0=0,\quad
a_1=2\left(\frac{1}{\rm Re}+\frac{1}{\rm Rm}\right),\quad b_1=4n,&\nn\\
&a_2=2\left(\frac{\beta^2{\rm N}}{\rm Rm}-3\right)n^2
+\frac{4\beta {\rm N}}{\rm Rm}n&\nn\\
&+\frac{2}{\rm Rm}\left(\frac{1}{\rm Re }+(1-{2{\rm Rb}\beta^2}){\rm N}\right)+4(1+{\rm Ro})+\frac{a_1^2}{4}, \quad b_2=\frac{3}{4}a_1b_1,&\nn\\
&a_3= a_1\left(\frac{\beta^2{\rm N}}{\rm Rm}-3\right)n^2+2a_1\beta\frac{{\rm N}}{\rm Rm}n+\frac{a_1}{\rm Rm}\left(\frac{1}{\rm Re}+(1-2{\rm Rb}\beta^2){\rm N}\right)+\frac{8}{\rm Rm}(1+{\rm Ro}),&\nn\\
&b_3= 4\left(\frac{\beta^2\rm N}{\rm Rm}-1\right)n^3+2n\left(4(1+{\rm Ro})+\frac{a_1^2}{4}+\frac{2}{\rm Rm} \left(\frac{1}{\rm Re}+{\rm N}\right)\right)&\nn\\
&-8(1-n^2+\beta n(1+{\rm Rb}))\frac{\beta{\rm N}}{\rm Rm},&\nn\\
&a_4=\left(\frac{\beta^2\rm N}{\rm Rm}-1\right)^2n^4+\frac{4\beta\rm N}{\rm Rm}\left(\frac{\beta^2\rm N}{\rm Rm}-1\right)n^3&\nn\\
&+\left(\frac{2}{\rm Rm}\left(\frac{\beta^2\rm N}{\rm Rm}-1\right)\left({\rm N}{+}\frac{1}{\rm Re}{-}2(\beta^2(1{+}{\rm Rb}){\rm N}{-}(1{+}{\rm Ro}){\rm Rm})\right){+}\frac{4\rm N^2\beta^2}{\rm Rm^2}-\frac{a_1^2}{4}\right)n^2&\nn\\
&+\frac{4\rm N\beta}{{\rm Rm}^2}\left(\rm N+\frac{1}{\rm Re}-2(\beta^2(1+{\rm Rb})\rm N-(1+{\rm Ro}){\rm Rm})\right)n&\nn\\
&+\frac{1}{\rm Rm^2}\left(\left({\rm N}+\frac{1}{\rm Re}\right)^2-4{\rm N}\left(\frac{\beta^2 \rm Rb}{\rm Re}- \rm Ro  Rm\right)+4(1+\rm Ro)-4 {\rm N}^2\beta^2 (1+\rm Rb)\right),&\nn\\
&b_4=a_1\left(\frac{\beta^2{\rm N}}{\rm Rm}-1\right)n^3+2a_1\frac{\beta{\rm N}}{\rm Rm} n^2-\frac{4\beta{\rm N}}{\rm Rm}\left(\frac{2}{\rm Rm}+\left(\frac{1}{\rm Rm}-\frac{1}{\rm Re}\right){\rm Ro}\right)&\nn\\
&+\left(\frac{a_1}{\rm Rm}\left(\frac{1}{\rm Re}+(1+2\beta^2({\rm Ro}- {\rm Rb}) ){\rm N}\right)  -\frac{8(1+{\rm Ro})}{\rm \rm Rm}\left(\frac{\beta^2{\rm N}}{\rm Rm} -1\right)\right)n.&
\ea

\section{Connection to the work by \cite{KGD1966}}\label{appB}

 The background flow in the Taylor-Couette cell between two co-axial cylinders of infinite length that rotate about their common axis
is
\be{m4b} \Omega(R)=a+\frac{b}{R^2},
\ee
where
\be{m5}
a=\frac{\mu_{\Omega}-\hat \eta^2}{1-\hat \eta^2}\Omega_1,\quad
b=\frac{1-\mu_{\Omega}}{1-\hat \eta^2}R_1^2\Omega_1,\quad
\hat \eta=\frac{R_1}{R_2},\quad \mu_{\Omega}=\frac{\Omega_2}{\Omega_1},
\ee
and $R_1$ and $R_2$ and $\Omega_1$ and $\Omega_2$ are the radii and angular velocities of the inner and outer cylinder, respectively.
The centrifugal acceleration of the
background flow \rf{m4b} is compensated by the pressure gradient:
$R\Omega^2=\frac{1}{\rho}\frac{\partial
p_0}{\partial R}$.
In view of the definition \rf{hro}
\be{hro1}
a=\Omega(1+{\rm Ro}).
\ee

In case when the gap between the cylinders $d=R_2-R_1$ is small with respect to the radii, the linearized equations derived by \cite{KGD1966}  for the stability of the viscous Taylor-Couette flow of hydrodynamics with respect to non-axisymmetric disturbances have the form
\be{kgd1}
L(D^2-(k_zd)^2)u'=-(k_zd)^2T\Omega_l(x)v',\quad L v'=u',
\ee
where $D=d/dx$, $R=R_1+xd$, $d/dR=d^{-1}d/dx$,  $T=-4a\Omega_1d^4/\nu^2$, $u'=u2ad\delta/(\nu \Omega_1)$, $v'=v/(R_1 \Omega_1)$, $\delta=d/R_1$,
\be{kgd2}
L=D^2-(k_zd)^2-i(\sigma+k\sqrt{T}~\Omega_l(x)),\quad \Omega_l(x)=1-(1-\mu_{\Omega})x,\quad x\in[0,1],
\ee
$\sigma=\omega d^2/\nu$, and $k=m\sqrt{-\frac{\Omega_1}{4a}}$.

For any $w \sim \exp(ik_R R)$, we have
\ba{kgd3}
Lw&=&-\frac{d^2}{\nu}\left(\nu(k_R^2+k_z^2) +i(\omega + m \Omega_1 \Omega_l)\right)w\nn\\
&=&-\frac{d^2}{\nu}\left(\nu|\boldsymbol{ k} |^2 +i(\omega + m \Omega_1 \Omega_l)\right)w=-\frac{d^2}{\nu}\left(\omega_{\nu} +i(\omega + m \Omega_1 \Omega_l)\right)w.
\ea
Consequently, if $u,v \sim \exp(ik_R R)$, the first equation in \rf{kgd1} becomes
\ba{kgd4}
-\frac{d^2}{\nu}\left(\omega_{\nu} +i(\omega + m \Omega_1 \Omega_l)\right)(-d^2|\boldsymbol{ k} |^2)\frac{2ad\delta}{\nu \Omega_1}u=-k_z^2 d^2(-4a\Omega_1d^4)\frac{\Omega_l }{\nu^2R_1 \Omega_1}v.
\ea
Simplifying this equation yields
\be{appb1}
\left(\omega_{\nu} +i(\omega + m \Omega_1 \Omega_l)\right)|\boldsymbol{ k} |^2u=v k_z^2 2\Omega_1\Omega_l \frac{d}{\delta R_1},
\ee
and, finally,
\be{appb2}
\left(\omega_{\nu} +i(\omega + m \Omega_1 \Omega_l)\right)u=2 \alpha^2 \Omega_1\Omega_l v,
\ee
where $\alpha=k_z/|\boldsymbol{ k}|$.
Analogously, the second equation in \rf{kgd1} takes the form
\be{appb3}
\left(\omega_{\nu} +i(\omega + m \Omega_1 \Omega_l)\right)v=-2a u\frac{\delta R_1}{d},
\ee
and, finally,
\be{appb4}
\left(\omega_{\nu} +i(\omega + m \Omega_1 \Omega_l)\right)v=-2\Omega(1+{\rm Ro}) u.
\ee
The equations \rf{appb2} and \rf{appb4} constitute the following system
\ba{kgd5}
\left(\omega_{\nu} +i(\omega + m \Omega_1 \Omega_l)\right)u&=&2 \alpha^2 \Omega_1\Omega_l v,\nn\\
\left(\omega_{\nu} +i(\omega + m \Omega_1 \Omega_l)\right)v&=&-2\Omega(1+{\rm Ro}) u.
\ea
Since $\Omega \approx \Omega_1 \Omega_l$, then
\ba{kgd6}
\left(\omega_{\nu} +i(\omega + m \Omega)\right)u&=&2 \alpha^2 \Omega v,\nn\\
\left(\omega_{\nu} +i(\omega + m \Omega)\right)v&=&-2\Omega(1+{\rm Ro}) u.
\ea
The system \rf{kgd6} can be written in the matrix form ${\bf A}{\bf w}=i\omega {\bf w}$ with the vector ${\bf w}=(u,v)^T$ and
\be{matrA}
{\bf A}=\left(
    \begin{array}{cc}
       -i m \Omega-\omega_{\nu} & 2 \alpha^2 \Omega \\
      -2\Omega(1+{\rm Ro}) & -i m \Omega-\omega_{\nu} \\
    \end{array}
  \right).
\ee
The matrix $\bf A$ is nothing else but the submatrix of the matrix $\bf H$ defined by Eq.~\rf{mri11} situated at the corner corresponding to the first two rows and first two columns of $\bf H$.

Let ${\bf A}_0={\bf A}(\omega_{\nu}=0)$, then

\be{hams1}
{\bf A}_0=-{\bf G}^{-1}\overline{\bf A}^T_0{\bf G},\quad {\bf G}=i{\bf J},\quad {\bf J}=\left(
                                                 \begin{array}{cc}
                                                   0 & -1 \\
                                                   1 & 0 \\
                                                 \end{array}
                                               \right),
\ee
meaning that ${\bf F}=-i{\bf G}{\bf A}_0=\overline{\bf F}^T$ is a Hermitian matrix:
\be{hams2}
{\bf F}=\left(
    \begin{array}{cc}
      2\Omega(1+{\rm Ro}) & im\Omega \\
      -im\Omega & 2\alpha^2\Omega \\
    \end{array}
  \right).
\ee
Then, the eigenvalue problem ${\bf A}_0{\bf w}=i\omega {\bf w}$ acquires the Hamiltonian form
\be{hams3}
{\bf Fw}=i\omega{\bf J}{\bf w},\quad {\bf F}=\overline{\bf F}^T
\ee
and has the eigenvalues
$
\omega_{1,2}=-m\Omega\pm 2\alpha\Omega\sqrt{1+{\rm Ro}}
$, corresponding to Kelvin waves.

\bibliographystyle{jfm}


\end{document}